\documentclass[aps,pra,twocolumn,superscriptaddress,showpacs]{revtex4-1}

\usepackage{graphicx}
\usepackage{amsmath}
\usepackage{comment}
\usepackage[colorlinks,linkcolor=blue,citecolor=blue]{hyperref}

\newcommand{\eqn}[1]{
\begin{eqnarray}
	#1
\end{eqnarray}
}

\begin{document}

\title{Multi-particle quantum dynamics under real-time observation}

\author{Yuto Ashida}
\affiliation{Department of Physics, University of Tokyo, 7-3-1 Hongo, Bunkyo-ku, Tokyo
113-0033, Japan}
\author{Masahito Ueda}
\affiliation{Department of Physics, University of Tokyo, 7-3-1 Hongo, Bunkyo-ku, Tokyo
113-0033, Japan}
\affiliation{RIKEN Center for Emergent Matter Science (CEMS), Wako, Saitama 351-0198, Japan
}

\date{\today}

\begin{abstract} 
Recent developments in quantum gas microscopy open up the possibility of real-time observation of quantum many-body systems. To understand the dynamics of atoms under such circumstances, we formulate the dynamics  under a real-time spatially resolved measurement and show that, in an appropriate limit of weak spatial resolution and strong atom-light coupling, the measurement indistinguishability of particles results in complete suppression of relative positional decoherence. As a consequence, quantum correlation in the multi-particle dynamics persists under a minimally destructive observation. We numerically demonstrate this for ultracold atoms in an optical lattice. Our theoretical framework can be applied to feedback control of quantum many-body systems which may be realized in subwavelength-spacing lattice systems. 
\end{abstract}


\maketitle


\section{Introduction}

A quantum system subject to continuous observation dramatically alters its dynamics due to the backaction of the measurement \cite{WHM10,GC07}.  An outstanding challenge is to understand the interplay between the quantum many-body dynamics and the measurement backaction in a continuously  monitored system. 
Experimental advances in ultracold atoms have brought about a unique opportunity to address this fundamental question. In fact, recent realizations of single-site resolved detection \cite{BWS09,SJF10,MM15,CLW15,PMF15,EH15} and addressing \cite{WC11,PMP15} of atoms in an optical lattice offer a powerful tool to investigate the quantum many-body dynamics at the single  particle level \cite{BWS10,EM11,FT13,FT132,FT15,SP15,IR15}. Further developments of {\it in-situ} imaging techniques will allow us to perform a nondestructive, real-time monitoring of the quantum many-body dynamics \cite{GN09,PYS14,PYS142,PPM152,YA15,YA16,MG15,AA15,WP16}. However, to achieve such a minimally destructive observation of a quantum gas, it is essential to perform a weak and continuous measurement, in contrast to the conventional strong (projective) and single-shot measurement \cite{BI12}. This type of continuous monitoring is a crucial step to apply measurement-based feedback control \cite{WHM93,SC11} to quantum many-body systems and will provide a new resource to quantum technology. Furthermore, such a monitoring may offer an opportunity to explore an effect of measurement backaction on quantum critical phenomena \cite{ND10,BF13,MV15,YAc16,YAc162}. 
Thus, it is necessary to develop a theoretical tool to analyze the quantum many-body dynamics under real-time observation. Of particular interest is a physical consequence of the indistinguishability of measurement signals in such a continuously monitored system. 

\begin{figure}[b]
\includegraphics[width=86mm]{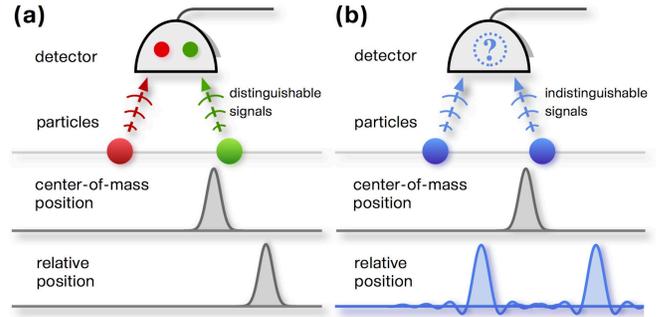}
\caption{\label{fig1} 
(color online). Schematic illustration of (a) distinguishable and (b) indistinguishable signals. If the signals coming from multiple particles cannot be distinguished, the relative positional coherence  in the multi-particle dynamics can be preserved
and a superposition of more than one relative position is allowed [(b) bottom]. If the signals are distinguishable, the relative position can take on a single value at each instant of time [(a) bottom]. On the other hand, the center-of-mass position behaves the same way for both cases.
}
\end{figure}

In this paper, we consider the quantum dynamics of multiple atoms in an optical lattice subject to a real-time spatially resolved observation, and obtain a stochastic many-body Schr\"{o}dinger equation in an appropriate limit of weak spatial resolution and strong atom-light coupling. Taking such a limit is essential to achieve a minimally destructive monitoring of quantum gases, as detailed below. While a general theory of the stochastic Schr\"{o}dinger equation has been well established \cite{BA95,BA98}, the derivation of the fundamental time-evolution equation for indistinguishable particles subject to weak continuous monitoring has long remained elusive \cite{BA13}. We show that the indistinguishability of measurable signals leads to the complete suppression of the relative positional decoherence, which makes a striking contrast with the case of distinguishable particles. This suppression induces the unique quantum transport dynamics that depends strongly on the measurement distinguishability of particles as schematically illustrated in Fig.~\ref{fig1}. In particular, the quantum correlation of indistinguishable particles is shown to persist under a minimally destructive observation. We demonstrate these results by numerical simulations of correlated quantum walks. In previous works concerning the site-resolved position measurement \cite{PH10,SK12,PD12,PD13,YY14} and  continuous position measurement of a single quantum particle \cite{CC87,DL88,DL882,BVP92,GMJ93}, the indistinguishability does not play such a nontrivial role. Our finding has the direct relevance to current techniques of {\it in}-{\it situ} observation of quantum gases \cite{BWS09,SJF10,MM15,CLW15,PMF15,EH15,GN09,PYS14,PYS142,PPM152,YA15,YA16,MG15,AA15,WP16}.

Our results also have a practical implication in preserving coherence in quantum technology and  opening up the possibility of performing measurement-based feedback control of quantum many-body systems. In particular, our theoretical description and finding may have important applications to recently proposed subwavelength-spacing lattices \cite{GM12,RIO13,GT15,SN15} as discussed below.

This paper is organized as follows. In Sec.~\ref{sec2}, we introduce a stochastic Schr\"{o}dinger equation as a model to describe the dynamics of particles trapped in a lattice subject to a spatially resolved continuous monitoring. In Sec.~\ref{sec3}, we consider a situation in which a spatial resolution is so low that we can continuously monitor the dynamics of atoms without substantial heating, making a sharp contrast to the conventional projective and single-shot measurement. To do so, we take the limit of weak spatial resolution and strong atom-light coupling and derive the stochastic time-evolution equation including the Wiener process for both indistinguishable and distinguishable particles. In Sec.~\ref{sec4}, we numerically demonstrate that the indistinguishability of multiple particles qualitatively alters their dynamics as a result of the absence of decoherence in relative positions. In Sec.~\ref{sec5}, we make several remarks on the relevance of our work to previous works and also on experimental conditions to achieve a minimally destructive observation. In Sec.~\ref{sec6}, we conclude this paper.

\begin{figure}[t]
\includegraphics[width=50mm]{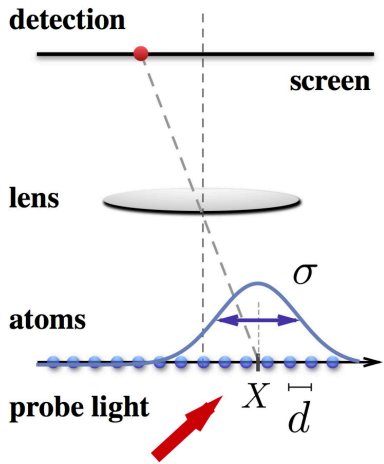}
\caption{\label{fig2} 
(color online). Schematic geometry of the system. Atoms trapped in an optical lattice (bottom, blue dots) are illuminated by an off-resonant light. A scattered photon is diffracted through a lens and detected on the screen. This induces the measurement backaction that causes the reduction of the wavefunction according to the outcome $X$ and a spatial resolution $\sigma$.
}
\end{figure}

\section{The Model\label{sec2}}
\subsection{Measurement protocol}
We consider $N$ atoms trapped in a lattice subject to a spatially resolved measurement as described in Ref. \cite{YA15}. For the sake of self-containedness, we briefly review the measurement protocol under consideration. 

Trapped atoms are described by  the Bose-Hubbard Hamiltonian $\hat{H}=-J\sum_{m}(\hat{b}^{\dagger}_{m}\hat{b}_{m+1}+{\rm H.c.})+(U/2)\sum_{m}\hat{n}_m(\hat{n}_m-1)$, where $J$ and $U$ are the hopping amplitude and the on-site interaction, respectively, $\hat{b}_{m}^{\dagger}$ is the creation operator of a boson at site $m$, and $\hat{n}_{m}=\hat{b}^{\dagger}_{m}\hat{b}_{m}$. We consider the eigenspace of the total particle number $\sum_{m}\hat{n}_{m}=N$. We measure atoms by illuminating an off-resonant probe light, and each scattered photon is diffracted through a lens aperture and detected on a screen as schematically illustrated in Fig.~\ref{fig2}. The detection causes the measurement backaction which induces the reduction of the many-body wavefunction of atoms, and constitutes a position measurement of identical particles \cite{YA15}. By integrating out the Heisenberg equation of motion and by using the fact that the length scale of the optical geometry is much larger than the lattice constant $d$, we obtain the positive frequency component of the scattered light field which gives the following measurement operator
\eqn{
\hat{M}(X)=\sqrt{\gamma}\sum_{m}f(X-md)\hat{b}_{m}^{\dagger}\hat{b}_{m},
} 
where $X$ denotes a measurement outcome, $\gamma$ is the parameter characterizing the strength of the measurement, and $f$ is an amplitude of a general point spread function. Here we assume that  $f$ is normalized as $\int f^{2}(X)dX=1$ and is symmetric about the origin $f(X)=f(-X)$. In the low-resolution limit discussed below, physical properties do not depend on the details of the point spread function, but only on an effective spatial resolution $\sigma$ (see Fig.~\ref{fig2}) which is introduced as an expansion coefficient of the displaced integration of $f$:
\eqn{\label{res}
\int dX f\left(X-md\right)f\left(X-ld\right)\simeq 1-\frac{(m-l)^2 d^2}{4\sigma^2}.
}
For example, if we consider a Gaussian point spread function $f(X)=e^{-X^2/(2\sigma^2)}/(\sigma^2\pi)^{1/4}$, the standard deviation serves as the effective resolution.

\subsection{Dynamics of atoms under continuous observation}
Let us formulate the dynamics of atoms subject to the continuous spatially resolved measurement discussed above. While we focus on the lattice system in this paper, the following discussions can be generalized to continuous space. When we consider photodetection as a measurement process, the quantum jump process associated with measurement backaction can be modeled by a marked point stochastic process \cite{HPB02}.
Thus, we describe the quantum dynamics under continuous monitoring by the following stochastic many-body Schr\"{o}dinger equation~(see Refs. \cite{BA95,BA98} for a general theory):
\eqn{\label{SSE1}
d|\psi\rangle&=&-\frac{i}{\hbar}\hat{H}|\psi\rangle dt\nonumber\\
&-&\frac{1}{2}\int dX\Bigl(\hat{M}^{\dagger}(X)\hat{M}(X)-\langle\hat{M}^{\dagger}(X)\hat{M}(X)\rangle\Bigr)|\psi\rangle dt  \nonumber\\
&+&\int dX\left(\frac{\hat{M}(X)|\psi\rangle}{\sqrt{\langle\hat{M}^{\dagger}(X)\hat{M}(X)\rangle}}-|\psi\rangle\right)dN(X;t).
}
The first term on the right-hand side describes a unitary evolution under the Bose-Hubbard Hamiltonian, while other terms describe a non-unitary time evolution caused by the measurement backaction. Here $\langle \cdots\rangle$ denotes an expectation value with respect to state $|\psi\rangle$. In the no-count process in which no photons are detected, the time evolution of the quantum state is governed by the second line in Eq.~(\ref{SSE1}). For a single-particle case, this part reduces to the unitary evolution because the non-Hermitian term $-1/2\int dX\hat{M}^{\dagger}(X)\hat{M}(X)$ is proportional to the identity operator and it contributes merely to a global constant factor. In contrast, for multiple indistinguishable particles, the time evolution during the no-count process is qualitatively different because quantum interference between indistinguishable particles alters the subsequent measurement rate depending on the atomic configuration. The third line in Eq.~(\ref{SSE1}) describes the quantum jump process associated with photodetections, where $dN(X;t)$ is a marked point process \cite{BA12} satisfying $dN(X;t)dN(Y;t)=\delta(X-Y)dN(X;t)$ and $E[dN(X;t)]=\langle\hat{M}^{\dagger}(X)\hat{M}(X)\rangle dt$. Here $E[\cdots]$ denotes an expectation value of the intensity of the stochastic process, which should be interpreted as the conditional mean given the past history. We note that $dN(X;t)$ is not a simple Poisson process since its intensity depends on a stochastic vector $|\psi\rangle$.

\section{Suppression of relative positional decoherence\label{sec3}}
\subsection{Minimally destructive observation}
Since our aim is to discuss real-time monitoring of a quantum gas, we consider a weak and continuous measurement instead of the conventional projective and single-shot measurements that entail  substantial heating and loss of atoms \cite{BI12,GM12,RIO13,GT15,SN15}. To achieve such a minimally destructive observation, it is essential to take the limit of both weak spatial resolution ($\sigma\gg d$) and strong atom-light coupling ($\gamma\gg J/\hbar$) while keeping the ratio $\gamma/\sigma^2$ finite. This limiting procedure is essential to obtain a stochastic Schr\"{o}dinger equation involving a nontrivial competition between the measurement backaction and the intrinsic unitary dynamics of the system  \cite{WHM10,CC87,DL88,DL882,BVP92,GMJ93}; otherwise, the unitary part of the evolution would be completely suppressed due to the quantum Zeno effect \cite{PYS142}, or the contribution from the measurement backaction would vanish. 

We begin by considering the total number of photocounts observed during a time interval $[0,t]$, which is $\tilde{N}(t)=\int N(t;X)dX$.
In the limit of weak spatial resolution, the mean of its rate of change is given by
\eqn{
\left\langle\frac{d\tilde{N}(t)}{dt}\right\rangle&=&\int dX\langle\hat{M}^{\dagger}(X)\hat{M}(X)\rangle\nonumber\\
&\simeq&\gamma\sum_{m,l}\left(1-\frac{(m-l)^2 d^2}{4\sigma^2}\right)\langle\hat{n}_{m}\hat{n}_{l}\rangle\simeq \gamma N^2,\nonumber\\
}
where we use Eq.~(\ref{res}) in deriving the second (approximate) equality. Since the number of counts within a given time interval goes to infinity in the limit $\gamma\to\infty$, from the central limit theorem, we can approximate the fluctuation of stochastic intensity  as 
\eqn{\label{wn1}
\frac{d\tilde{N}(t)-\int dX\langle\hat{M}^{\dagger}(X)\hat{M}(X)\rangle dt}{N\sqrt{\gamma}}\simeq dW(t),
}
where $dW(t)$ is the Wiener stochastic process satisfying $E[dW(t)]=0$ and $\left(dW(t)\right)^2=dt$. Similarly, as for the spatial dependent stochastic process $dN(X;t)$, we approximate its fluctuation as
\eqn{\label{wn2}
\frac{dN(X;t)-\langle\hat{M}^{\dagger}(X)\hat{M}(X)\rangle dt}{\sqrt{\langle\hat{M}^{\dagger}(X)\hat{M}(X)\rangle}}\simeq dW(X;t),
}
where $dW(X;t)$ is the position-dependent Wiener process satisfying $E[dW(X;t)]=0$ and $dW(X;t)dW(Y;t)=\delta(X-Y)dt$.

\subsection{Time-evolution equation for indistinguishable particles}
Let us discuss the time-evolution equation that governs the dynamics of indistinguishable particles under real-time observation described above. To do so, we consider the time-evolution equation of the density matrix $\hat{\rho}=|\psi\rangle\langle\psi|$ and then substitute Eqs.~(\ref{wn1}) and (\ref{wn2}) into it to take the limit of weak spatial resolution and strong atom-light coupling outlined above. In particular, from the condition of a weak spatial resolution, we assume that atoms are not spatially resolved, i.e., the spatial extension of the atomic wavefunction is less than the resolution $\sigma$. The technical details of calculations are given in Appendix \ref{app1}. We thus obtain the following stochastic many-body Schr\"{o}dinger equation including the center-of-mass operator $\hat{X}_{CM}=\sum_{m}m\hat{n}_{m}/N$ in the backaction terms:

\eqn{\label{SSE3}
d\hat{\rho}=&-&\frac{i}{\hbar}\left[\hat{H},\hat{\rho}\right]dt-\frac{N^2\gamma d^2}{4\sigma^2}\left[\hat{X}_{CM},\left[\hat{X}_{CM},\hat{\rho}\right]\right]dt \nonumber\\
&+&\sqrt{\frac{N^2\gamma d^2}{2\sigma^2}}\left\{\hat{X}_{CM}-\langle\hat{X}_{CM}\rangle,\hat{\rho}\right\}dW(t),
}
where $dW(t)$ is the Wiener stochastic process and originates from an overall fluctuation in the observed signals (see Appendix \ref{app1}).
Remarkably, in contrast to distinguishable particles \cite{DL88,DL89}, the relative positional decoherence term is absent, which has profound implications for quantum transport as demonstrated below. The crucial observation here is that the indistinguishability of measurable signals leads to the  interference of the amplitudes between particles at  different sites, resulting in the cancellation of the relative positional decoherence. In this respect, we note that the suppression can also occur for intrinsically distinguishable yet  practically indistinguishable particles in the actual measurement process (for example, imagine the coherent light scattering by isotopes).  In practice, the  measurement distinguishability would be relevant when one considers, for example, the state-selective imaging \cite{WC11,FT13,FT132,FT15,SP15,PPM152} or the polarization measurement \cite{DJS10}.

We note that the suppression of the relative positional decoherence can also be interpreted as an emergence of a  decoherence-free subspace (DFS) \cite{BA00,FP02}; the continuous measurement backaction can generate an effective DFS which is given by a certain value of the center-of-mass coordinate of the many-body system. Such an emergence of a DFS is a manifestation of the symmetry property in the measurement operator acting on indistinguishable particles. 

The time-evolution equation~(\ref{SSE3}) shows the first generalization of a model of continuous position measurement  \cite{CC87,DL88,DL882,BVP92,GMJ93} to quantum many-body systems. Due to the difficulty of generalizing the original derivations  \cite{CC87,DL88,DL882,BVP92,GMJ93}, a continuous position measurement model for indistinguishable particles has long remained elusive \cite{BA13} because photons scattered at two different sites can be distinguished. The measurement indistinguishability could not play a nontrivial role for the cases of site-resolved measurement \cite{PH10,SK12,PD12,PD13,YY14} and single particle models \cite{CC87,DL88,DL882,BVP92,GMJ93}. Note that we do {\it not} assume that the degrees of freedom of relative positions are frozen unlike in a rigid system \cite{DL89,GGC86,GGC90}. 


\subsection{\label{dissec}Time-evolution equation for distinguishable particles}
Let us next discuss the time-evolution equation for distinguishable particles. We consider $N$ distinguishable particles and introduce the associated measurement operators $\hat{M}_{i}(X)$ given by
\eqn{\label{Sp0}
\hat{M}_{i}(X)=\sqrt{\gamma_{i}}\sum_{x}f(X-xd)|x\rangle_{i}{}_{i}\langle x|,
}
where $i$ is the label of particles, $\gamma_{i}$ is a detection rate of measurable signals of particle $i$, $f$ denotes a general point spread function, and $x$ is the discretized position of a lattice. Then, the stochastic Schr\"{o}dinger equation for distinguishable particles becomes
\eqn{\label{Sp1}
d|\psi\rangle&=&-\frac{i}{\hbar}\hat{H}|\psi\rangle dt\nonumber\\
&-&\frac{1}{2}\sum_{i=1}^{N}\int dX_{i}\Bigl(\hat{M}_{i}^{\dagger}(X_{i})\hat{M}_{i}(X_{i})\nonumber\\
&&\;\;\;\;\;\;\;\;\;\;\;\;\;\;\;\;\;\;\;\;\;\;\;-\langle\hat{M}_{i}^{\dagger}(X_{i})\hat{M}_{i}(X_{i})\rangle\Bigr)|\psi\rangle dt  \nonumber\\
&+&\sum_{i=1}^{N}\!\int \!dX_{i}\!\left(\!\frac{\hat{M}_{i}(X_{i})|\psi\rangle}{\sqrt{\langle\hat{M}_{i}^{\dagger}(X_{i})\hat{M}_{i}(X_{i})\rangle}}\!-\!|\psi\rangle\!\right)\!dN_{i}(X_{i};t),\nonumber\\
}
where $dN_{i}(X;t)$ are the stochastic processes associated with a measurable signal of particle $i$ and obey $dN_{i}(X;t)dN_{j}(X;t)=\delta_{ij}\delta(X-Y)dN_{i}(X;t)$. For non-interacting distinguishable particles, the Hamiltonian takes the form $\hat{H}=-J\sum_{i,m}(\hat{b}_{i,m}^{\dagger}\hat{b}_{i,m+1}+{\rm H.c.})$, where we assume that the hopping rate is independent of particle species $i$. For simplicity, we assume that the  particles are not entangled in the initial state. By taking the limit of weak spatial resolution and strong coupling of Eq.~(\ref{Sp1}) (see Appendix \ref{app2}), we can obtain the following time-evolution equation for distinguishable particles:

\eqn{\label{Sp2pre}
d\hat{\rho}=&-&\frac{i}{\hbar}\left[\hat{H},\hat{\rho}\right]dt-\sum_{i=1}^{N}\frac{\gamma_{i}d^2}{4\sigma^2}\left[\hat{x}_{i},\left[\hat{x}_{i},\hat{\rho}\right]\right]dt\nonumber\\
&+&\sum_{i=1}^{N}\sqrt{\frac{\gamma_{i}d^2}{2\sigma^2}}\left\{\hat{x}_{i}-\langle\hat{x}_{i}\rangle,\hat{\rho}\right\}dW_{i}(t),
}
where we introduce the position operator of each particle, $\hat{x}_{i}=\sum_{m}m|m\rangle_{i}{}_{i}\langle m|$, and $dW_i(t)$'s are independent Wiener processes satisfying $E[dW_i(t)]=0$ and $dW_{i}(t)dW_{j}(t)=\delta_{i,j}dt$. Physically, the Wiener process $dW_i$ represents fluctuations of measureble signals coming from the $i$-th particle. 
For simplicity, we assume that the scattering rate $\gamma_{i}=\gamma$ is independent of $i$. Then, the time-evolution equation~(\ref{Sp2pre}) is simplified as follows:

\eqn{\label{Sp2}
d\hat{\rho}=&-&\frac{i}{\hbar}\left[\hat{H},\hat{\rho}\right]dt-\frac{N\gamma d^2}{4\sigma^2}\left[\hat{X}_{CM},\left[\hat{X}_{CM},\hat{\rho}\right]\right]dt\nonumber\\
&-&\frac{\gamma d^2}{4\sigma^2}\sum_{i=1}^{N}\left[\hat{r}_{i},\left[\hat{r}_{i},\hat{\rho}\right]\right]dt \nonumber\\
&+&\sqrt{\frac{N\gamma d^2}{2\sigma^2}}\left\{\hat{X}_{CM}-\langle\hat{X}_{CM}\rangle,\hat{\rho}\right\}dW(t)\nonumber\\
&+&\sqrt{\frac{\gamma d^2}{2\sigma^2}}\sum_{i=1}^{N}\left\{\hat{r}_{i}-\langle\hat{r}_{i}\rangle,\hat{\rho}\right\} dW_{i}(t),
}
where we define the center-of-mass operator $\hat{X}_{CM}=\sum_{i=1}^{N}\hat{x}_{i}/N$, the relative coordinate, $\hat{r}_{i}=\hat{x}_{i}-\hat{X}_{CM}$, and $dW(t)\equiv\sqrt{1/N}\sum_{i}dW_{i}(t)$. We note that $dW(t)$ is not independent from $dW_{i}(t)$. Equations~(\ref{Sp2pre}) and (\ref{Sp2}) are consistent with the known result \cite{DL88,DL89} for a model of continuous position measurement of distinguishable particles, which can be obtained by applying a single-particle model \cite{CC87,DL88,DL882,BVP92,GMJ93}. 

The key point is that, for distinguishable particles, there exists the contribution from relative positional decoherence (the third and the fifth terms on the right-hand side of Eq.~(\ref{Sp2})). Since the Wiener processes in the time-evolution equations originate from the noise in the observed signals, the additional dissipative terms in Eq.~(\ref{Sp2}) result from the distinguishability of the particles by the different observable signals.  Such a contribution can be completely suppressed for indistinguishable particles as shown in the previous section and significantly change the behavior of quantum transport as discussed in the next section.

\subsection{Emergence of three temporal regimes}
The interplay between the measurement distinguishability and the quantum transport can result in a unique multi-particle dynamics. To discuss such a physical consequence of the absence of the relative positional decoherence for indistinguishable particles, let us focus on the decoherence rate of the off-diagonal term $\langle\{n_m\}|E[\hat{\rho}]|\{n'_m\}\rangle$ of the density matrix:
\eqn{\label{Dec1}
\Gamma_{\{n_{m}\},\{n'_{m}\}}=\frac{N^2\gamma d^2}{4\sigma^2}\left(X_{CM}-X'_{CM}\right)^2.
}
This can be obtained by considering matrix elements of the second term on the right-hand side of Eq.~(\ref{SSE3}). Here $\{n_m\}$ denotes the Fock state and $X_{CM}$ ($X'_{CM}$) is the center-of-mass coordinate of the state $\{n_m\}$ ($\{n'_m\}$). From Eq.~(\ref{Dec1}), we can infer the following distinct regimes in the time evolution:
\\
{\it (i) Center-of-mass collapse regime.} A superposition of different center-of-mass states rapidly decoheres. Consequently, the many-body wavefunction collapses into a state whose center-of-mass coordinate takes a well-localized value. This collapse occurs on the time scale of about $4\sigma^2/(N^2\gamma d^2L^2)$, where $L$ denotes a typical distance between the center-of-mass positions of the superposed Fock states. 
\\
{\it (ii) Inertial regime.} Once the center-of-mass coordinate is well localized, the coherence within the subspace of the Fock states that take on the values close to the center of mass is preserved due to the absence of the relative positional decoherence. Thus, the relative quantum motion of multiple indistinguishable particles is basically unaffected by the continuous monitoring. 
\\
{\it (iii) Diffusive regime.} In the long-time limit ($t\gg 4\sigma^2/(\gamma d^2)$), the coherence between the nearest Fock states whose center-of-mass coordinates differ by the minimal distance $\delta X_{CM}=1/N$ is eventually lost, and particles start to undergo random walk. While such an eventual diffusive transport can also be seen for distinguishable particles,  the diffusion constant is qualitatively different depending on quantum statistics (see Fig.~\ref{fig4}(a)).

\section{Numerical Simulations for two particles\label{sec4}}
\begin{figure}[b]
\includegraphics[width=86mm]{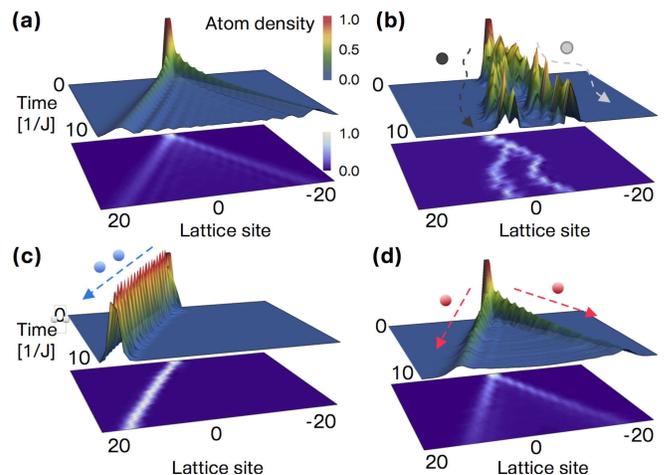}
\caption{\label{fig3} 
(color online). Real-time quantum dynamics of two particles under continuous observation. (a) Unitary time evolution. Without measurement backaction, quantum statistics hardly affects the atomic density. (b-d) Quantum dynamics subject to the measurement backaction for (b) distinguishable particles, (c) bosons, and (d) fermions, calculated for $\Gamma=2.0$. Owing to the absence of the relative positional decoherence, (c) bunched and (d) anti-correlated  quantum walks persist, whereas distinguishable particles exhibit uncorrelated random walks (b).
}
\end{figure}
To demonstrate the general properties discussed in the preceding section, we perform numerical simulations at the single-trajectory level on the basis of Eq.~(\ref{SSE3}) for indistinguishable particles and Eq.~(\ref{Sp2}) for distinguishable particles. A particularly interesting and important regime is the inertial regime where the coherent quantum dynamics of relative motion persists owing to the absence of relative positional decoherence. Such robustness can manifest itself in quantum walks of two particles. From now on, we set $U=0$ and focus on the quantum behavior caused by purely quantum statistics and measurement indistinguishability of particles. We note that, for fermions, the operators $\hat{b}_{m}^{\dagger}$ and $\hat{b}_{m}$ in the Hamiltonian $\hat{H}$ should be interpreted as the fermionic operators. As long as we consider non-interacting particles, qualitatively the same argument should apply to a larger number of particles. 
To make a fair comparison, we perform numerical simulations under the condition that the total detection rate $\Gamma$ of signals is the same.

While an expectation value of a physical observable can be obtained from the expectation value of the density matrix $E[\hat{\rho}]$ that is obtained by integrating out its master equation, the aim of our numerical simulations here is to reveal the real-time dynamics of quantum particles subject to the weak and continuous monitoring at the single-trajectory level. Indeed, there exists certain information, such as the widths of the localized wavepackets, that are inherent to such real-time dynamics and can be inferred only from considering the dynamics at the single-trajectory level as discussed below.
 
\subsection{Inertial regime}
We consider an initial condition in which two particles are localized at adjacent sites. Without measurement backaction, the density profile is almost independent of quantum statistics of the particles involved (Fig.~\ref{fig3}a) \cite{LY12}. In this case, the peaks of the atom density propagate along two straight lines, indicating ballistic transport of  atoms: $\langle x^{2}\rangle=2J^{2}t^{2}/\hbar^{2}$ \cite{MG98}. In contrast, with  measurement backaction, the quantum dynamics dramatically changes depending on quantum statistics. While distinguishable particles exhibit uncorrelated random walks (Fig.~\ref{fig3}b), indistinguishable particles exhibit ballistic and strongly correlated walks (Fig.~\ref{fig3}c, d). 
Ballistic transport implies that quantum coherence between Fock states within a well-localized center-of-mass subspace is preserved, which is  characteristic of the inertial regime. This results from the fact that indistinguishability of measurable signals protects the quantum system from relative positional decoherence. The strong correlation arises from the multi-particle interference between quantum amplitudes of identical particles. 

For bosons, two atoms move in the same direction due to constructive interference of identical bosons (Fig.~\ref{fig3}c).  Because of rapid collapse of the center-of-mass coordinate, two bosons form a localized wave packet. The collapse occurs on the time scale of about $t_{\rm col}\simeq(3\sqrt{2}\hbar^{2}/(\Gamma J^2))^{1/3}$. In practice, to experimentally explore such localization of the bosonic wavefunction, it is advantageous to verify the absence of two-particle interference, since the difference from the coherent time evolution of Fig.~\ref{fig3}(a) is hardly distinguished with a density profile after taking the ensemble average over measurement outcomes. One can also confirm this boson bunching by coincidence measurement.

In contrast, for fermions, two atoms move in the opposite directions due to destructive interference (Fig.~\ref{fig3}d), and the center-of-mass coordinate takes a localized value around zero due to the anti-correlation, so that the weak continuous monitoring does not appreciably alter the quantum dynamics compared with the coherent dynamics of Fig.~\ref{fig3}(a), making striking contrast with the site-resolved measurement \cite{SK12}. 

\subsection{Diffusive regime}
\begin{figure}[b]
\includegraphics[width=86mm]{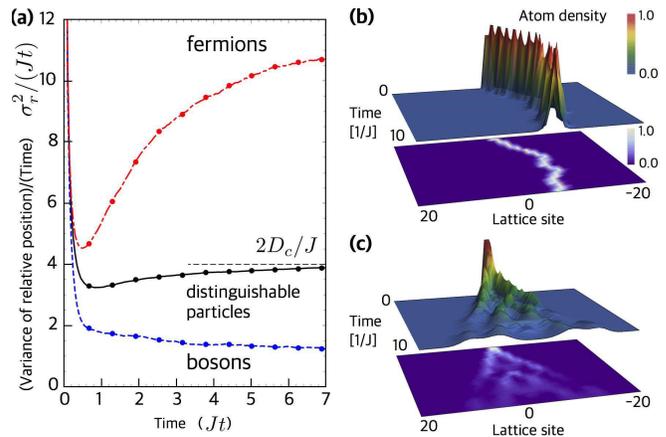}
\caption{\label{fig4} 
(color online). Diffusive regime of two-particle quantum dynamics under continuous observation. (a) Variance of the relative distance divided by the elapsed time. Distinguishable particles (solid curve) show exponential convergence to $2D_{c}/J$ (dashed line). Bosons (blue dashed curve) and fermions (red dash-dotted curve) exhibit a stronger diffusion and a weaker diffusion, respectively, than distinguishable particles. A typical realization of the atomic density is shown for (b) bosons and (c) fermions. The results are calculated with $\Gamma=16.0$ and averaged over $10^{3}$ realizations. 
}
\end{figure}

To discuss the diffusive regime, we numerically simulate the dynamics for a larger strength of the measurement $\Gamma$.  Figure \ref{fig4} (a) plots the square $\sigma_{r}^{2}$ of relative distance between two particles averaged over many realizations. For distinguishable particles, relative motion  eventually exhibits random-walk behavior with the diffusion constant $D_{c}=16J^{2}\sigma^{2}/(\gamma\hbar^{2} d^2)$ (see App. \ref{app3}). In contrast, quantum statistics serves as an effective attractive (repulsive) interaction for bosons (fermions), resulting in a diffusion constant smaller (larger) than $D_{c}$. Figure \ref{fig4} (b) and (c) show typical realizations of two-particle transport for bosons and fermions, respectively. Here we do see that quantum statistics can affect the diffusive transport because the coherence between Fock states that have the same center-of-mass value remains nonvanishing owing to the complete suppression of relative positional decoherence.

\section{Discussions\label{sec5}}
The time-evolution equation~(\ref{SSE1}) implies that, since relevant decoherence operators now reduce to the center-of-mass coordinate alone, we can generalize the theory of quantum feedback control \cite{WHM93,SC11} that was originally developed in homodyne detection in quantum optics to quantum many-body systems. The center-of-mass position should be controllable, for example, by displacing the entire optical lattice. 

In the opposite high-resolution limit $\sigma\ll d$, Eq.~(\ref{SSE1}) reproduces the Lindblad-type equation: $E[\dot{\hat{\rho}}]=-(i/\hbar)[\hat{H},\hat{\rho}]-(\gamma/2)\sum_{m}[\hat{n}_m,[\hat{n}_m,E[\hat{\rho}]]]$, which has been used in the study of the quantum dynamics under the site-resolved measurement \cite{PH10,PD12,PD13,YY14}. In the absence of interaction $U=0$, the coherence is rapidly lost during the time scale of $\sim 1/\gamma$ and particles exhibit diffusive behavior characteristic of classical random walk \cite{ES72,BTA03,SH10,SA11}. In the presence of nonzero interaction $U\neq0$, an exotic behavior such as anomalous diffusion can be found \cite{PD12,PD13}. Thus, exploring such a nontrivial role of the interaction in the weak continuous measurement regime merits further study. In particular, Eq.~(\ref{SSE3}) enables us to investigate a real-time dynamics of a many-body system conditioned on measurement outcomes.

While we consider here an ideal situation of a unit collection efficiency of signals, our theory can easily be generalized to take into account uncollected signals by using the standard treatment of open quantum systems \cite{WHM10,KJ06}. For a possible effect of heating, although such effects can usually be made negligible, for example, by conducting an experiment in the deep Lamb-Dicke regime \cite{JE03} or using Raman sideband cooling \cite{PYS14,PYS142,CLW15,PMF15}, continuous imaging may cause heating of trapped atoms in the long-time regime. In such a case, the theory should be modified by including higher bands \cite{PH10}. To achieve the real-time monitoring discussed in this paper, one needs to perform a low-resolved imaging of optical lattice systems. While the resolution $\sigma$ can be changed by adjusting a numerical aperture of lens \cite{WC112}, such a control also affects the collection efficiency of signals and may lead to a higher heating rate of atoms. To meet the requirement, we suggest that recently proposed subwavelength lattices \cite{GM12,RIO13,GT15,SN15} will offer candidate systems for implementing such a minimally destructive imaging. This is because, in those systems, the lattice constant can be made much shorter than that of optical lattice systems, and both a low spatial resolution and a high collection efficiency of photons can be attained simultaneously. Another possible scheme is to use imaging light whose wavelength is longer than the lattice constant in which a heating effect can be substantially suppressed.
\section{Conclusion\label{sec6}}
In this paper, we have constructed a theoretical framework for real-time monitoring of the multi-particle dynamics in an optical lattice. 
To discuss a real-time, minimally destructive observation of quantum gases, we consider the low-resolved and continuous position measurement of atoms. In the limit of weak spatial resolution and strong atom-light coupling,  we have shown that the indistinguishability of particles in a measurement process protects the system from the relative positional decoherence, leading to the unique transport dynamics. We show that the interplay between the suppression of the relative positional decoherence and quantum transport results in the distinct regimes in the time evolution. In particular, we find the regime where quantum correlation in the dynamics of indistinguishable particles persists owing to the suppression of relative positional decoherence. In the long-time regime, particles start a random walk behavior, where the diffusion constant strongly depends on particle species. We numerically demonstrate these results by quantum walks of two atoms trapped in an optical lattice.
Our findings should be investigated by sub-wavelength lattice systems or using an imaging light whose wavelength is longer than the lattice constant.
Such real-time observation will be applicable for realizing quantum feedback control of quantum many-body systems. Also, it may provide an interesting opportunity to explore an influence of measurement back-action on quantum critical phenomena \cite{ND10,BF13,MV15,YAc16,YAc162}. 

\begin{acknowledgments}
We acknowledge Y. Kuramochi, T. Fukuhara, S. Furukawa, T. Shitara, R. Hamazaki, K. Saito, and A. Alberti for fruitful discussions. This work was supported by KAKENHI Grant No. JP26287088 from the Japan Society for the Promotion of Science, and a Grant-in-Aid for Scientific Research on Innovative Areas ``Topological Materials Science" (KAKENHI Grant No. JP15H05855), and the Photon Frontier Network Program from MEXT of Japan, ImPACT Program of Council for Science, Technology and Innovation (Cabinet Office, Government of Japan), and the Mitsubishi Foundation. Y. A. acknowledges support from JSPS (Grant No. JP16J03613). 
\end{acknowledgments}

\appendix
\section{\label{app1}Limit of weak spatial resolution and strong atom-light coupling}
Here we take the weak-resolution and strong atom-light coupling limit of Eq.~(\ref{SSE1}) and derive the time-evolution equation~(\ref{SSE3}) that describes the dynamics of indistinguishable particles under real-time observation. Using Eq.~(\ref{SSE1}) and the multiplication rule of $dN(X;t)$, we obtain the time-evolution equation of $\hat{\rho}=|\psi\rangle\langle\psi|$ as follows:
\eqn{\label{SSE}
d\hat{\rho}&=&\!-\frac{i}{\hbar}\left[\hat{H},\hat{\rho}\right]\!\!+\!{\cal{D}}[\hat{\rho}]dt\!+\!\int \!dX\!\!\left(\!\frac{\hat{M}(X)\hat{\rho}\hat{M}^{\dagger}(X)}{\langle\hat{M}^{\dagger}(X)\hat{M}(X)\rangle}\!\!-\!\!\hat{\rho}\right)\nonumber\\
&&\;\;\;\times\left(dN(X;t)-\langle\hat{M}^{\dagger}(X)\hat{M}(X)\rangle\right),\\
{\cal{D}}[\hat{\rho}]&=&\!\int \!dX\left(\hat{M}(X)\hat{\rho}\hat{M}^{\dagger}(X)\!-\!\frac{1}{2}\left\{\hat{M}^{\dagger}(X)\hat{M}(X),\hat{\rho}\right\}\right).\nonumber\\ \label{dissip}
}
For the sake of concreteness, we consider the operator with the Gaussian point-spread function: $\hat{M}(X)=\sqrt{\gamma/\sqrt{\pi\sigma^{2}}}\sum_{m}\exp\left[-(X-md)^2/(2\sigma^2)\right]\hat{n}_{m}$. Then, the dissipator ${\cal{D}}[\hat{\rho}]$ in Eq.~(\ref{dissip}) can be calculated as
\eqn{
{\cal D}[\hat{\rho}]&\simeq& -\frac{\gamma}{2}\sum_{m,l}\left(1-\frac{(m-l)^2 d^2}{4\sigma^2}\left[\hat{n}_{l},\left[\hat{n}_{m},\hat{\rho}\right]\right]\right)\nonumber\\
&=&-\frac{N^2\gamma d^2}{4\sigma^2}\left[\hat{X}_{CM},\left[\hat{X}_{CM},\hat{\rho}\right]\right],\label{deridis}
}
where we assume that the spatial resolution of the measurement is so low that interference peaks of atoms in a cluster cannot be resolved. Technically, this is equivalent to requiring that matrix elements of the operator $\hat{n}_{m}\hat{n}_{l}\hat{\rho}+\hat{\rho}\hat{n}_{l}\hat{n}_{m}-2\hat{n}_{m}\hat{\rho}\hat{n}_{l}$ rapidly vanish for $|m-l|>\sigma/d$ so that the expansion of the exponential function is justified \cite{BA13}. In deriving the last equality, we use the particle conservation $\sum_{m}\hat{n}_{m}=N\hat{I}$, where $N$ is the number of atoms and $\hat{I}$ is the identity operator, and introduce the center-of-mass operator $\hat{X}_{CM}=\sum_{m}\hat{n}_{m}/N$. Equation~(\ref{deridis}) gives the first line in the time-evolution equation~(\ref{SSE3}).

Next, we consider the weak spatial-resolution and strong atom-light coupling limit of the fluctuating contribution that is the last term in Eq.~(\ref{SSE}). To complete the derivation of Eq.~(\ref{SSE3}), the following contribution has to be shown to go to zero in this limit: 
\eqn{
&&\int dX \,R(X;t)
\,\frac{dN(X;t)-\langle \hat M^\dagger(X)\hat M(X)\rangle dt}{\sqrt{\langle \hat M^\dagger(X)\hat M(X)\rangle}}\nonumber\\
&\simeq&\int dX\, R(X;t)dW(X;t),\label{RW}
}
where we use the approximation (\ref{wn2}) and introduce
\eqn{
&&R(X;t)=\sqrt{\langle \hat M^\dagger(X)\hat M(X)\rangle}\nonumber\\
&&\times\!\left(\frac{\hat M(X)
\hat\rho
\hat M^\dagger(X)}{\langle \hat M^\dagger(X)\hat M(X)\rangle}\!-\!\hat\rho\!
-\!\frac { d}{\sqrt{2}\, \sigma}\left\{\hat X_{CM}\!-\!\langle \hat X_{CM}\rangle,
\hat \rho\right\}\right).\nonumber\\\label{Rcalc}
}
To do so, we expand the Gaussian function in the operator $\hat{M}(X)$ as 
\eqn{
\sum_{m}e^{-\frac{(X-md)^2}{2\sigma^2}}\hat{n}_{m}&\simeq& e^{-\frac{X^2}{2\sigma^2}}\sum_{m}\left(1+\frac{mXd}{\sigma^2}\right)\hat{n}_{m}\label{expansionderi} \\
&=&e^{-\frac{X^2}{2\sigma^2}}N\left(\hat{I}+\frac{Xd}{\sigma^2}\hat{X}_{CM}\right).\label{expderi2}
}
Use of the approximation~(\ref{expansionderi}) in Eq.~(\ref{RW}) can be justified as follows. First, since we consider the situation in which atoms cannot spatially be resolved, in particular, we assume that atoms are positioned around the site at $m=0$ and a size of cluster of the particles is less than the spatial resolution $\sigma$, as numerically investigated above. Equivalently, we assume that matrix elements of the operator $\hat{n}_{m}\hat{\rho}\hat{n}_{l}$ rapidly vanish when $m>\sigma/d$ or $l>\sigma/d$. We thus neglect higher-order terms of $m$ in Eq.~(\ref{expansionderi}). Second, we also neglect higher-order terms of $X$ since they result in higher-order terms in $d/\sigma$ after the integration over $X$. Specifically, we can show that, after performing the integration in Eq.~(\ref{RW}), $O(X^n d^n/\sigma^{2n})$ contributions ultimately provide $O(\sqrt{\sigma}\cdot(d/\sigma)^n)$ contributions (as inferred from, e.g., Eq.~(\ref{wienerf}) below). Such higher-order terms can be neglected compared with the leading contribution (\ref{expderi2}) in the limit of a weak spatial resolution $d/\sigma\to 0$.
Thus, we calculate the first two terms in Eq.~(\ref{Rcalc}) as
\eqn{
&&\int_{-\infty}^{\infty}\!\!dX\!\!\left(\!\frac{\hat{M}(X)\hat{\rho}\hat{M}^{\dagger}(X)}{\sqrt{\langle\hat{M}^{\dagger}(X)\hat{M}(X)\rangle}}\!-\!\sqrt{\langle\hat{M}^{\dagger}(X)\hat{M}(X)\rangle}\hat{\rho}\!\right)\nonumber\\
&&\;\;\;\;\;\;\;\;\;\;\;\;\;\;\;\;\;\;\;\;\;\;\;\;\;\;\;\;\;\;\;\;\;\;\;\;\;\;\;\;\;\;\;\;\;\;\;\;\;\;\;\;\;\;\;\;\;\;\;\;\;\;\;\;\times dW(X;t)\nonumber\\
&\simeq&\!\!\sqrt{\frac{N^2\gamma d^2}{\sqrt{\pi}\sigma^5}}\!\int_{-\infty}^{\infty}\!dXdW(\!X;t)Xe^{-\frac{X^2}{2\sigma^2}}\left\{\!\hat{X}_{CM}\!-\!\langle\hat{X}_{CM}\rangle,\hat{\rho}\right\}.\nonumber\\
&=&\sqrt{\frac{N^2\gamma d^2}{2\sigma^2}}\left\{\hat{X}_{CM}-\langle\hat{X}_{CM}\rangle,\hat{\rho}\right\}dW(t),\label{SSE2}
}
where we use the fact that a linear superposition of Wiener stochastic processes is a Wiener process aside from a constant factor:
\eqn{\label{wienerf}
\int_{-\infty}^{\infty}dXdW(X;t)Xe^{-\frac{X^2}{2\sigma^2}}=\sqrt{\frac{\sigma^3 \sqrt{\pi}}{2}}dW(t).
}
Here $dW$ is the standard Wiener process satisfying $\left(dW(t)\right)^2=dt$ and $E[dW(t)]=0$. Equation~(\ref{SSE2}) cancels out the contribution resulting from the last term in Eq.~(\ref{Rcalc}), which means that Eq.~(\ref{RW}) goes to zero in the limit of weak spatial resolution, which completes the derivation of Eq.~(\ref{SSE3}). 
 
We note that if we consider a single particle, Eq.~(\ref{SSE3}) reproduces the previous results \cite{CC87,DL88,DL882,BVP92,GMJ93} for the continuous position measurement model of a single particle. We also note that, by employing the effective spatial resolution defined by Eq.~(\ref{res}), one can derive the weak spatial-resolution and strong atom-light coupling limit for a general point spread function along the same line as in the above calculations.

\section{\label{app2}Derivation of the time-evolution equation for distinguishable particles}
We here derive the time-evolution equation~(\ref{Sp2pre}) for distinguishable particles from Eq.~(\ref{Sp1}). Using the multiplication rules of $dN_{i}(X;t)$, we obtain the time-evolution equation of $\hat{\rho}=|\psi\rangle\langle\psi|$:
\eqn{\label{SSEdis}
d\hat{\rho}&=&-\frac{i}{\hbar}\left[\hat{H},\hat{\rho}\right]\nonumber\\
&&+\sum_{i=1}^{N}\Biggl[{\cal{D}}_{i}[\hat{\rho}]dt+\!\int \!dX_{i}\left(\frac{\hat{M}_{i}(X_{i})\hat{\rho}\hat{M}_{i}^{\dagger}(X_{i})}{\langle\hat{M}_{i}^{\dagger}(X_{i})\hat{M}_{i}(X_{i})\rangle}-\hat{\rho}\right)\nonumber\\
&&\;\;\;\;\;\;\;\;\times\left(dN_{i}(X_{i};t)-\langle\hat{M}_{i}^{\dagger}(X_{i})\hat{M}_{i}(X_{i})\rangle\right)\Biggr],\\
{\cal{D}}_{i}[\hat{\rho}]&=&\!\int \!dX\left(\hat{M}_{i}(X)\hat{\rho}\hat{M}^{\dagger}_{i}(X)\!-\!\frac{1}{2}\left\{\hat{M}_{i}^{\dagger}(X)\hat{M}_{i}(X),\hat{\rho}\right\}\right).\nonumber\\\label{dissipdis}
}
As in the previous section, the dissipator ${\cal D}_{i}[\hat{\rho}]$ can be calculated in the weak resolution limit as
\eqn{\label{dissderi}
{\cal D}_{i}[\hat{\rho}]\simeq-\frac{\gamma_{i}d^2}{4\sigma^2}\left[\hat{x}_{i},\left[\hat{x}_{i},\hat{\rho}\right]\right],
}
where we introduce the position operator of each particle, $\hat{x}_{i}=\sum_{m}m|m\rangle_{i}{}_{i}\langle m|$, and use $\hat{x}_{i}^{2}=\sum_{m}m^2|m\rangle_{i}{}_{i}\langle m|$, ${}_{i}\langle m|l\rangle_{i}=\delta_{m,l}$, and $\sum_{m}|m\rangle_{i}{}_{i}\langle m|=\hat{I}$. Equation~(\ref{dissderi}) gives the first line in Eq.~(\ref{Sp2pre}). 

We next consider the remaining terms in Eq.~(\ref{SSEdis}). To complete the derivation of Eq.~(\ref{Sp2pre}), the following contribution has to be shown to go to zero: 
\eqn{
&&\int dX \,R_{i}(X;t)
\,\frac{dN_{i}(X;t)-\langle \hat M_{i}^\dagger(X)\hat M_{i}(X)\rangle dt}{\sqrt{\langle \hat M_{i}^\dagger(X)\hat M_{i}(X)\rangle}}\nonumber\\
&\simeq&\int dX\, R_{i}(X;t)dW_{i}(X;t),\label{RWdis}
}
where we use the central limit theorem and introduce
\eqn{
R_{i}(X;t)&=&\sqrt{\langle \hat M_{i}^\dagger(X)\hat M_{i}(X)\rangle}\nonumber\\
&\times&\!\left(\frac{\hat M_{i}(X)
\hat\rho
\hat M_{i}^\dagger(X)}{\langle \hat M_{i}^\dagger(X)\hat M_{i}(X)\rangle}\!-\!\hat\rho\!
-\!\frac { d}{\sqrt{2}\, \sigma}\left\{\hat x_{i}\!-\!\langle \hat x_{i}\rangle,\hat \rho\right\}\right).\nonumber\\\label{Rcalcdis}
}
Since we assume that the particles are not entangled in the initial state and are non-interacting as described in Sec.~\ref{dissec}, the density matrix $\hat{\rho}$ can be decomposed into the direct product of $\hat{\rho}_{i}$ for each particle $i$. We can thus show   
the convergence of Eq.~(\ref{RWdis}) by applying the derivation  (\ref{SSE2}) in the previous section to a single-particle case $N=1$.

\section{\label{app3}Derivation of the diffusion constant for distinguishable particles}
We here derive an analytical formula of the diffusion constant of the relative distance of distinguishable particles. To this end, we first consider a single-particle model. We denote $\langle n|\hat{\rho}|m\rangle=\rho_{n,m}$, where $|n\rangle$ is the state in which a particle localizes at site $n$. Then, the non-selective time evolution is described by
\eqn{\label{dec2}
\dot{\rho}_{n,m}&=&iJ\bigl(\rho_{n+1,m}+\rho_{n-1,m}-\rho_{n,m+1}-\rho_{n,m-1}\bigr)\nonumber\\
&&-\frac{\Gamma}{4}(n-m)^2\rho_{n,m}.
}
We are interested in the diffusive regime, i.e., in the long-time limit. In this regime, as can be seen from Eq.~(\ref{dec2}), the coherence between remote different sites should be neglected because the decoherence rate becomes larger as the distance between different sites increases. Hence, we only consider the coherence between neighboring sites $\rho_{n,n\pm1}$ and also assume that the time evolution of these values can be neglected $\dot{\rho}_{n,n\pm 1}\simeq 0$ in the stationary regime. This results in the following expressions
\eqn{
\rho_{n,n\pm1}&\simeq& \frac{4iJ}{\Gamma}\left(\rho_{n\pm1,n\pm1}-\rho_{n,n}\right),\nonumber\\
\rho_{n\pm1,n}&\simeq& -\frac{4iJ}{\Gamma}\left(\rho_{n\pm1,n\pm1}-\rho_{n,n}\right).
}
By substituting these into the time evolution equation of the on-site density component $\rho_{n,n}$, we obtain 
\eqn{
\dot{\rho}_{n,n}&=&iJ\left(\rho_{n+1,n}+\rho_{n-1,n}-\rho_{n,n+1}-\rho_{n,n-1}\right)\nonumber\\
&\simeq&\frac{8J^2}{\Gamma}\left(\rho_{n+1,n+1}-2\rho_{n,n}+\rho_{n-1,n-1}\right),
}
which is the diffusion equation with the diffusion constant $8J^2/\Gamma$. We note that similar discussions for the site-resolved decoherence model can be found in Refs. \cite{ES72,SH10}. 

For non-interacting distinguishable particles, the variance of the relative distance between two particles is given as $\sigma_{r}^{2}=\langle(\hat{x}_{1}-\hat{x}_{2})^2\rangle=\langle \hat{x}_{1}^2 \rangle+\langle \hat{x}_{2}^2\rangle-2\langle\hat{x}_{1}\rangle\langle\hat{x}_{2}\rangle$, where $\hat{x}_{1(2)}$ is the position operator of particle 1 (2). If particles are initially localized in single sites, the expectation values of the positions $\langle\hat{x}_{1,2}\rangle$ should remain in the initial values. In particular, we choose the initial condition of two adjacent particles $\langle\hat{x}_{1}\rangle_{0}=0$ and $\langle\hat{x}_{2}\rangle_{0}=1$ as in the numerical simulations. As a result, we obtain the long-time limit behavior of $\sigma_{r}$ as
\eqn{
\sigma_{r}^2=\frac{32J^2t}{\Gamma}\equiv 2D_{c}t,
} 
where we introduce the diffusion constant $D_{c}$ for the relative distance.

\bibliography{reference}

\begin{thebibliography}{68}%
\makeatletter
\providecommand \@ifxundefined [1]{%
 \@ifx{#1\undefined}
}%
\providecommand \@ifnum [1]{%
 \ifnum #1\expandafter \@firstoftwo
 \else \expandafter \@secondoftwo
 \fi
}%
\providecommand \@ifx [1]{%
 \ifx #1\expandafter \@firstoftwo
 \else \expandafter \@secondoftwo
 \fi
}%
\providecommand \natexlab [1]{#1}%
\providecommand \enquote  [1]{``#1''}%
\providecommand \bibnamefont  [1]{#1}%
\providecommand \bibfnamefont [1]{#1}%
\providecommand \citenamefont [1]{#1}%
\providecommand \href@noop [0]{\@secondoftwo}%
\providecommand \href [0]{\begingroup \@sanitize@url \@href}%
\providecommand \@href[1]{\@@startlink{#1}\@@href}%
\providecommand \@@href[1]{\endgroup#1\@@endlink}%
\providecommand \@sanitize@url [0]{\catcode `\\12\catcode `\$12\catcode
  `\&12\catcode `\#12\catcode `\^12\catcode `\_12\catcode `\%12\relax}%
\providecommand \@@startlink[1]{}%
\providecommand \@@endlink[0]{}%
\providecommand \url  [0]{\begingroup\@sanitize@url \@url }%
\providecommand \@url [1]{\endgroup\@href {#1}{\urlprefix }}%
\providecommand \urlprefix  [0]{URL }%
\providecommand \Eprint [0]{\href }%
\providecommand \doibase [0]{http://dx.doi.org/}%
\providecommand \selectlanguage [0]{\@gobble}%
\providecommand \bibinfo  [0]{\@secondoftwo}%
\providecommand \bibfield  [0]{\@secondoftwo}%
\providecommand \translation [1]{[#1]}%
\providecommand \BibitemOpen [0]{}%
\providecommand \bibitemStop [0]{}%
\providecommand \bibitemNoStop [0]{.\EOS\space}%
\providecommand \EOS [0]{\spacefactor3000\relax}%
\providecommand \BibitemShut  [1]{\csname bibitem#1\endcsname}%
\let\auto@bib@innerbib\@empty
\bibitem [{\citenamefont {Wiseman}\ and\ \citenamefont
  {Milburn}(2010)}]{WHM10}%
  \BibitemOpen
  \bibfield  {author} {\bibinfo {author} {\bibfnamefont {H.}~\bibnamefont
  {Wiseman}}\ and\ \bibinfo {author} {\bibfnamefont {G.}~\bibnamefont
  {Milburn}},\ }\href@noop {} {\emph {\bibinfo {title} {Quantum Measurement and
  Control}}}\ (\bibinfo  {publisher} {Cambridge University Press},\ \bibinfo
  {year} {2010})\BibitemShut {NoStop}%
\bibitem [{\citenamefont {Guerlin}\ \emph {et~al.}(2007)\citenamefont
  {Guerlin}, \citenamefont {Bernu}, \citenamefont {Del\`eglise}, \citenamefont
  {Sayrin}, \citenamefont {Gleyzes}, \citenamefont {Kuhr}, \citenamefont
  {Brune}, \citenamefont {Raimond},\ and\ \citenamefont {Haroche}}]{GC07}%
  \BibitemOpen
  \bibfield  {author} {\bibinfo {author} {\bibfnamefont {C.}~\bibnamefont
  {Guerlin}}, \bibinfo {author} {\bibfnamefont {J.}~\bibnamefont {Bernu}},
  \bibinfo {author} {\bibfnamefont {S.}~\bibnamefont {Del\`eglise}}, \bibinfo
  {author} {\bibfnamefont {C.}~\bibnamefont {Sayrin}}, \bibinfo {author}
  {\bibfnamefont {S.}~\bibnamefont {Gleyzes}}, \bibinfo {author} {\bibfnamefont
  {S.}~\bibnamefont {Kuhr}}, \bibinfo {author} {\bibfnamefont {M.}~\bibnamefont
  {Brune}}, \bibinfo {author} {\bibfnamefont {J.-M.}\ \bibnamefont {Raimond}},
  \ and\ \bibinfo {author} {\bibfnamefont {S.}~\bibnamefont {Haroche}},\ }\href
  {http://www.nature.com/nature/journal/v448/n7156/full/nature06057.html}
  {\bibfield  {journal} {\bibinfo  {journal} {Nature}\ }\textbf {\bibinfo
  {volume} {448}},\ \bibinfo {pages} {889} (\bibinfo {year}
  {2007})}\BibitemShut {NoStop}%
\bibitem [{\citenamefont {Bakr}\ \emph {et~al.}(2009)\citenamefont {Bakr},
  \citenamefont {Gillen}, \citenamefont {Peng}, \citenamefont {F{\"o}lling},\
  and\ \citenamefont {Greiner}}]{BWS09}%
  \BibitemOpen
  \bibfield  {author} {\bibinfo {author} {\bibfnamefont {W.~S.}\ \bibnamefont
  {Bakr}}, \bibinfo {author} {\bibfnamefont {J.~I.}\ \bibnamefont {Gillen}},
  \bibinfo {author} {\bibfnamefont {A.}~\bibnamefont {Peng}}, \bibinfo {author}
  {\bibfnamefont {S.}~\bibnamefont {F{\"o}lling}}, \ and\ \bibinfo {author}
  {\bibfnamefont {M.}~\bibnamefont {Greiner}},\ }\href
  {http://www.nature.com/nature/journal/v462/n7269/full/nature08482.html}
  {\bibfield  {journal} {\bibinfo  {journal} {Nature}\ }\textbf {\bibinfo
  {volume} {462}},\ \bibinfo {pages} {74} (\bibinfo {year} {2009})}\BibitemShut
  {NoStop}%
\bibitem [{\citenamefont {Sherson}\ \emph {et~al.}(2010)\citenamefont
  {Sherson}, \citenamefont {Weitenberg}, \citenamefont {Endres}, \citenamefont
  {Cheneau}, \citenamefont {Bloch},\ and\ \citenamefont {Kuhr}}]{SJF10}%
  \BibitemOpen
  \bibfield  {author} {\bibinfo {author} {\bibfnamefont {J.~F.}\ \bibnamefont
  {Sherson}}, \bibinfo {author} {\bibfnamefont {C.}~\bibnamefont {Weitenberg}},
  \bibinfo {author} {\bibfnamefont {M.}~\bibnamefont {Endres}}, \bibinfo
  {author} {\bibfnamefont {M.}~\bibnamefont {Cheneau}}, \bibinfo {author}
  {\bibfnamefont {I.}~\bibnamefont {Bloch}}, \ and\ \bibinfo {author}
  {\bibfnamefont {S.}~\bibnamefont {Kuhr}},\ }\href
  {http://www.nature.com/nature/journal/v467/n7311/full/nature09378.html}
  {\bibfield  {journal} {\bibinfo  {journal} {Nature}\ }\textbf {\bibinfo
  {volume} {467}},\ \bibinfo {pages} {68} (\bibinfo {year} {2010})}\BibitemShut
  {NoStop}%
\bibitem [{\citenamefont {Miranda}\ \emph {et~al.}(2015)\citenamefont
  {Miranda}, \citenamefont {Inoue}, \citenamefont {Okuyama}, \citenamefont
  {Nakamoto},\ and\ \citenamefont {Kozuma}}]{MM15}%
  \BibitemOpen
  \bibfield  {author} {\bibinfo {author} {\bibfnamefont {M.}~\bibnamefont
  {Miranda}}, \bibinfo {author} {\bibfnamefont {R.}~\bibnamefont {Inoue}},
  \bibinfo {author} {\bibfnamefont {Y.}~\bibnamefont {Okuyama}}, \bibinfo
  {author} {\bibfnamefont {A.}~\bibnamefont {Nakamoto}}, \ and\ \bibinfo
  {author} {\bibfnamefont {M.}~\bibnamefont {Kozuma}},\ }\href {\doibase
  10.1103/PhysRevA.91.063414} {\bibfield  {journal} {\bibinfo  {journal} {Phys.
  Rev. A}\ }\textbf {\bibinfo {volume} {91}},\ \bibinfo {pages} {063414}
  (\bibinfo {year} {2015})}\BibitemShut {NoStop}%
\bibitem [{\citenamefont {Cheuk}\ \emph {et~al.}(2015)\citenamefont {Cheuk},
  \citenamefont {Nichols}, \citenamefont {Okan}, \citenamefont {Gersdorf},
  \citenamefont {Ramasesh}, \citenamefont {Bakr}, \citenamefont {Lompe},\ and\
  \citenamefont {Zwierlein}}]{CLW15}%
  \BibitemOpen
  \bibfield  {author} {\bibinfo {author} {\bibfnamefont {L.~W.}\ \bibnamefont
  {Cheuk}}, \bibinfo {author} {\bibfnamefont {M.~A.}\ \bibnamefont {Nichols}},
  \bibinfo {author} {\bibfnamefont {M.}~\bibnamefont {Okan}}, \bibinfo {author}
  {\bibfnamefont {T.}~\bibnamefont {Gersdorf}}, \bibinfo {author}
  {\bibfnamefont {V.~V.}\ \bibnamefont {Ramasesh}}, \bibinfo {author}
  {\bibfnamefont {W.~S.}\ \bibnamefont {Bakr}}, \bibinfo {author}
  {\bibfnamefont {T.}~\bibnamefont {Lompe}}, \ and\ \bibinfo {author}
  {\bibfnamefont {M.~W.}\ \bibnamefont {Zwierlein}},\ }\href {\doibase
  10.1103/PhysRevLett.114.193001} {\bibfield  {journal} {\bibinfo  {journal}
  {Phys. Rev. Lett.}\ }\textbf {\bibinfo {volume} {114}},\ \bibinfo {pages}
  {193001} (\bibinfo {year} {2015})}\BibitemShut {NoStop}%
\bibitem [{\citenamefont {Parsons}\ \emph {et~al.}(2015)\citenamefont
  {Parsons}, \citenamefont {Huber}, \citenamefont {Mazurenko}, \citenamefont
  {Chiu}, \citenamefont {Setiawan}, \citenamefont {Wooley-Brown}, \citenamefont
  {Blatt},\ and\ \citenamefont {Greiner}}]{PMF15}%
  \BibitemOpen
  \bibfield  {author} {\bibinfo {author} {\bibfnamefont {M.~F.}\ \bibnamefont
  {Parsons}}, \bibinfo {author} {\bibfnamefont {F.}~\bibnamefont {Huber}},
  \bibinfo {author} {\bibfnamefont {A.}~\bibnamefont {Mazurenko}}, \bibinfo
  {author} {\bibfnamefont {C.~S.}\ \bibnamefont {Chiu}}, \bibinfo {author}
  {\bibfnamefont {W.}~\bibnamefont {Setiawan}}, \bibinfo {author}
  {\bibfnamefont {K.}~\bibnamefont {Wooley-Brown}}, \bibinfo {author}
  {\bibfnamefont {S.}~\bibnamefont {Blatt}}, \ and\ \bibinfo {author}
  {\bibfnamefont {M.}~\bibnamefont {Greiner}},\ }\href {\doibase
  10.1103/PhysRevLett.114.213002} {\bibfield  {journal} {\bibinfo  {journal}
  {Phys. Rev. Lett.}\ }\textbf {\bibinfo {volume} {114}},\ \bibinfo {pages}
  {213002} (\bibinfo {year} {2015})}\BibitemShut {NoStop}%
\bibitem [{\citenamefont {Haller}\ \emph {et~al.}(2015)\citenamefont {Haller},
  \citenamefont {Hudson}, \citenamefont {Kelly}, \citenamefont {Cotta},
  \citenamefont {Bruno}, \citenamefont {Bruce},\ and\ \citenamefont
  {Kuhr}}]{EH15}%
  \BibitemOpen
  \bibfield  {author} {\bibinfo {author} {\bibfnamefont {E.}~\bibnamefont
  {Haller}}, \bibinfo {author} {\bibfnamefont {J.}~\bibnamefont {Hudson}},
  \bibinfo {author} {\bibfnamefont {A.}~\bibnamefont {Kelly}}, \bibinfo
  {author} {\bibfnamefont {D.~A.}\ \bibnamefont {Cotta}}, \bibinfo {author}
  {\bibfnamefont {P.}~\bibnamefont {Bruno}}, \bibinfo {author} {\bibfnamefont
  {G.~D.}\ \bibnamefont {Bruce}}, \ and\ \bibinfo {author} {\bibfnamefont
  {S.}~\bibnamefont {Kuhr}},\ }\href
  {http://www.nature.com/nphys/journal/vaop/ncurrent/abs/nphys3403.html}
  {\bibfield  {journal} {\bibinfo  {journal} {Nat. Phys.}\ }\textbf {\bibinfo
  {volume} {11}},\ \bibinfo {pages} {738} (\bibinfo {year} {2015})}\BibitemShut
  {NoStop}%
\bibitem [{\citenamefont {Weitenberg}\ \emph
  {et~al.}(2011{\natexlab{a}})\citenamefont {Weitenberg}, \citenamefont
  {Endres}, \citenamefont {Sherson}, \citenamefont {Cheneau}, \citenamefont
  {Schau\ss}, \citenamefont {Fukuhara}, \citenamefont {Bloch},\ and\
  \citenamefont {Kuhr}}]{WC11}%
  \BibitemOpen
  \bibfield  {author} {\bibinfo {author} {\bibfnamefont {C.}~\bibnamefont
  {Weitenberg}}, \bibinfo {author} {\bibfnamefont {M.}~\bibnamefont {Endres}},
  \bibinfo {author} {\bibfnamefont {J.~F.}\ \bibnamefont {Sherson}}, \bibinfo
  {author} {\bibfnamefont {M.}~\bibnamefont {Cheneau}}, \bibinfo {author}
  {\bibfnamefont {P.}~\bibnamefont {Schau\ss}}, \bibinfo {author}
  {\bibfnamefont {T.}~\bibnamefont {Fukuhara}}, \bibinfo {author}
  {\bibfnamefont {I.}~\bibnamefont {Bloch}}, \ and\ \bibinfo {author}
  {\bibfnamefont {S.}~\bibnamefont {Kuhr}},\ }\href
  {http://www.nature.com/nature/journal/v471/n7338/full/nature09827.html}
  {\bibfield  {journal} {\bibinfo  {journal} {Nature}\ }\textbf {\bibinfo
  {volume} {471}},\ \bibinfo {pages} {319} (\bibinfo {year}
  {2011}{\natexlab{a}})}\BibitemShut {NoStop}%
\bibitem [{\citenamefont {Preiss}\ \emph
  {et~al.}(2015{\natexlab{a}})\citenamefont {Preiss}, \citenamefont {Ma},
  \citenamefont {Tai}, \citenamefont {Lukin}, \citenamefont {Rispoli},
  \citenamefont {Zupancic}, \citenamefont {Lahini}, \citenamefont {Islam},\
  and\ \citenamefont {Greiner}}]{PMP15}%
  \BibitemOpen
  \bibfield  {author} {\bibinfo {author} {\bibfnamefont {P.~M.}\ \bibnamefont
  {Preiss}}, \bibinfo {author} {\bibfnamefont {R.}~\bibnamefont {Ma}}, \bibinfo
  {author} {\bibfnamefont {M.~E.}\ \bibnamefont {Tai}}, \bibinfo {author}
  {\bibfnamefont {A.}~\bibnamefont {Lukin}}, \bibinfo {author} {\bibfnamefont
  {M.}~\bibnamefont {Rispoli}}, \bibinfo {author} {\bibfnamefont
  {P.}~\bibnamefont {Zupancic}}, \bibinfo {author} {\bibfnamefont
  {Y.}~\bibnamefont {Lahini}}, \bibinfo {author} {\bibfnamefont
  {R.}~\bibnamefont {Islam}}, \ and\ \bibinfo {author} {\bibfnamefont
  {M.}~\bibnamefont {Greiner}},\ }\href {\doibase 10.1126/science.1260364}
  {\bibfield  {journal} {\bibinfo  {journal} {Science}\ }\textbf {\bibinfo
  {volume} {347}},\ \bibinfo {pages} {1229} (\bibinfo {year}
  {2015}{\natexlab{a}})}\BibitemShut {NoStop}%
\bibitem [{\citenamefont {Bakr}\ \emph {et~al.}(2010)\citenamefont {Bakr},
  \citenamefont {Peng}, \citenamefont {Tai}, \citenamefont {Ma}, \citenamefont
  {Simon}, \citenamefont {Gillen}, \citenamefont {F{\"o}lling}, \citenamefont
  {Pollet},\ and\ \citenamefont {Greiner}}]{BWS10}%
  \BibitemOpen
  \bibfield  {author} {\bibinfo {author} {\bibfnamefont {W.~S.}\ \bibnamefont
  {Bakr}}, \bibinfo {author} {\bibfnamefont {A.}~\bibnamefont {Peng}}, \bibinfo
  {author} {\bibfnamefont {M.~E.}\ \bibnamefont {Tai}}, \bibinfo {author}
  {\bibfnamefont {R.}~\bibnamefont {Ma}}, \bibinfo {author} {\bibfnamefont
  {J.}~\bibnamefont {Simon}}, \bibinfo {author} {\bibfnamefont {J.~I.}\
  \bibnamefont {Gillen}}, \bibinfo {author} {\bibfnamefont {S.}~\bibnamefont
  {F{\"o}lling}}, \bibinfo {author} {\bibfnamefont {L.}~\bibnamefont {Pollet}},
  \ and\ \bibinfo {author} {\bibfnamefont {M.}~\bibnamefont {Greiner}},\ }\href
  {\doibase 10.1126/science.1192368} {\bibfield  {journal} {\bibinfo  {journal}
  {Science}\ }\textbf {\bibinfo {volume} {329}},\ \bibinfo {pages} {547}
  (\bibinfo {year} {2010})}\BibitemShut {NoStop}%
\bibitem [{\citenamefont {{Endres}}\ \emph {et~al.}(2011)\citenamefont
  {{Endres}}, \citenamefont {{Cheneau}}, \citenamefont {{Fukuhara}},
  \citenamefont {{Weitenberg}}, \citenamefont {{Schau{\ss}}}, \citenamefont
  {{Gross}}, \citenamefont {{Mazza}}, \citenamefont {{Ba{\~n}uls}},
  \citenamefont {{Pollet}}, \citenamefont {{Bloch}},\ and\ \citenamefont
  {{Kuhr}}}]{EM11}%
  \BibitemOpen
  \bibfield  {author} {\bibinfo {author} {\bibfnamefont {M.}~\bibnamefont
  {{Endres}}}, \bibinfo {author} {\bibfnamefont {M.}~\bibnamefont {{Cheneau}}},
  \bibinfo {author} {\bibfnamefont {T.}~\bibnamefont {{Fukuhara}}}, \bibinfo
  {author} {\bibfnamefont {C.}~\bibnamefont {{Weitenberg}}}, \bibinfo {author}
  {\bibfnamefont {P.}~\bibnamefont {{Schau{\ss}}}}, \bibinfo {author}
  {\bibfnamefont {C.}~\bibnamefont {{Gross}}}, \bibinfo {author} {\bibfnamefont
  {L.}~\bibnamefont {{Mazza}}}, \bibinfo {author} {\bibfnamefont {M.~C.}\
  \bibnamefont {{Ba{\~n}uls}}}, \bibinfo {author} {\bibfnamefont
  {L.}~\bibnamefont {{Pollet}}}, \bibinfo {author} {\bibfnamefont
  {I.}~\bibnamefont {{Bloch}}}, \ and\ \bibinfo {author} {\bibfnamefont
  {S.}~\bibnamefont {{Kuhr}}},\ }\href
  {http://science.sciencemag.org/content/334/6053/200} {\bibfield  {journal}
  {\bibinfo  {journal} {Science}\ }\textbf {\bibinfo {volume} {334}},\ \bibinfo
  {pages} {200} (\bibinfo {year} {2011})}\BibitemShut {NoStop}%
\bibitem [{\citenamefont {{Fukuhara}}\ \emph {et~al.}(2013)\citenamefont
  {{Fukuhara}}, \citenamefont {{Kantian}}, \citenamefont {{Endres}},
  \citenamefont {{Cheneau}}, \citenamefont {{Schau{\ss}}}, \citenamefont
  {{Hild}}, \citenamefont {{Bellem}}, \citenamefont {{Schollw{\"o}ck}},
  \citenamefont {{Giamarchi}}, \citenamefont {{Gross}}, \citenamefont
  {{Bloch}},\ and\ \citenamefont {{Kuhr}}}]{FT13}%
  \BibitemOpen
  \bibfield  {author} {\bibinfo {author} {\bibfnamefont {T.}~\bibnamefont
  {{Fukuhara}}}, \bibinfo {author} {\bibfnamefont {A.}~\bibnamefont
  {{Kantian}}}, \bibinfo {author} {\bibfnamefont {M.}~\bibnamefont {{Endres}}},
  \bibinfo {author} {\bibfnamefont {M.}~\bibnamefont {{Cheneau}}}, \bibinfo
  {author} {\bibfnamefont {P.}~\bibnamefont {{Schau{\ss}}}}, \bibinfo {author}
  {\bibfnamefont {S.}~\bibnamefont {{Hild}}}, \bibinfo {author} {\bibfnamefont
  {D.}~\bibnamefont {{Bellem}}}, \bibinfo {author} {\bibfnamefont
  {U.}~\bibnamefont {{Schollw{\"o}ck}}}, \bibinfo {author} {\bibfnamefont
  {T.}~\bibnamefont {{Giamarchi}}}, \bibinfo {author} {\bibfnamefont
  {C.}~\bibnamefont {{Gross}}}, \bibinfo {author} {\bibfnamefont
  {I.}~\bibnamefont {{Bloch}}}, \ and\ \bibinfo {author} {\bibfnamefont
  {S.}~\bibnamefont {{Kuhr}}},\ }\href
  {http://www.nature.com/nphys/journal/v9/n4/full/nphys2561.html} {\bibfield
  {journal} {\bibinfo  {journal} {Nat. Phys.}\ }\textbf {\bibinfo {volume}
  {9}},\ \bibinfo {pages} {235} (\bibinfo {year} {2013})}\BibitemShut {NoStop}%
\bibitem [{\citenamefont {Fukuhara}\ \emph {et~al.}(2013)\citenamefont
  {Fukuhara}, \citenamefont {Schausz}, \citenamefont {Endres}, \citenamefont
  {Hild}, \citenamefont {Cheneau}, \citenamefont {Bloch},\ and\ \citenamefont
  {Gross}}]{FT132}%
  \BibitemOpen
  \bibfield  {author} {\bibinfo {author} {\bibfnamefont {T.}~\bibnamefont
  {Fukuhara}}, \bibinfo {author} {\bibfnamefont {P.}~\bibnamefont {Schausz}},
  \bibinfo {author} {\bibfnamefont {M.}~\bibnamefont {Endres}}, \bibinfo
  {author} {\bibfnamefont {S.}~\bibnamefont {Hild}}, \bibinfo {author}
  {\bibfnamefont {M.}~\bibnamefont {Cheneau}}, \bibinfo {author} {\bibfnamefont
  {I.}~\bibnamefont {Bloch}}, \ and\ \bibinfo {author} {\bibfnamefont
  {C.}~\bibnamefont {Gross}},\ }\href
  {http://www.nature.com/nature/journal/v502/n7469/full/nature12541.html}
  {\bibfield  {journal} {\bibinfo  {journal} {Nature}\ }\textbf {\bibinfo
  {volume} {502}},\ \bibinfo {pages} {76} (\bibinfo {year} {2013})}\BibitemShut
  {NoStop}%
\bibitem [{\citenamefont {Fukuhara}\ \emph {et~al.}(2015)\citenamefont
  {Fukuhara}, \citenamefont {Hild}, \citenamefont {Zeiher}, \citenamefont
  {Schau\ss{}}, \citenamefont {Bloch}, \citenamefont {Endres},\ and\
  \citenamefont {Gross}}]{FT15}%
  \BibitemOpen
  \bibfield  {author} {\bibinfo {author} {\bibfnamefont {T.}~\bibnamefont
  {Fukuhara}}, \bibinfo {author} {\bibfnamefont {S.}~\bibnamefont {Hild}},
  \bibinfo {author} {\bibfnamefont {J.}~\bibnamefont {Zeiher}}, \bibinfo
  {author} {\bibfnamefont {P.}~\bibnamefont {Schau\ss{}}}, \bibinfo {author}
  {\bibfnamefont {I.}~\bibnamefont {Bloch}}, \bibinfo {author} {\bibfnamefont
  {M.}~\bibnamefont {Endres}}, \ and\ \bibinfo {author} {\bibfnamefont
  {C.}~\bibnamefont {Gross}},\ }\href {\doibase 10.1103/PhysRevLett.115.035302}
  {\bibfield  {journal} {\bibinfo  {journal} {Phys. Rev. Lett.}\ }\textbf
  {\bibinfo {volume} {115}},\ \bibinfo {pages} {035302} (\bibinfo {year}
  {2015})}\BibitemShut {NoStop}%
\bibitem [{\citenamefont {Schau\ss{}}\ \emph {et~al.}(2015)\citenamefont
  {Schau\ss{}}, \citenamefont {Zeiher}, \citenamefont {Fukuhara}, \citenamefont
  {Hild}, \citenamefont {Cheneau}, \citenamefont {Macri}, \citenamefont {Pohl},
  \citenamefont {Bloch},\ and\ \citenamefont {Gross}}]{SP15}%
  \BibitemOpen
  \bibfield  {author} {\bibinfo {author} {\bibfnamefont {P.}~\bibnamefont
  {Schau\ss{}}}, \bibinfo {author} {\bibfnamefont {J.}~\bibnamefont {Zeiher}},
  \bibinfo {author} {\bibfnamefont {T.}~\bibnamefont {Fukuhara}}, \bibinfo
  {author} {\bibfnamefont {S.}~\bibnamefont {Hild}}, \bibinfo {author}
  {\bibfnamefont {M.}~\bibnamefont {Cheneau}}, \bibinfo {author} {\bibfnamefont
  {T.}~\bibnamefont {Macri}}, \bibinfo {author} {\bibfnamefont
  {T.}~\bibnamefont {Pohl}}, \bibinfo {author} {\bibfnamefont {I.}~\bibnamefont
  {Bloch}}, \ and\ \bibinfo {author} {\bibfnamefont {C.}~\bibnamefont
  {Gross}},\ }\href {\doibase 10.1126/science.1258351} {\bibfield  {journal}
  {\bibinfo  {journal} {Science}\ }\textbf {\bibinfo {volume} {347}},\ \bibinfo
  {pages} {1455} (\bibinfo {year} {2015})}\BibitemShut {NoStop}%
\bibitem [{\citenamefont {{Islam}}\ \emph {et~al.}(2015)\citenamefont
  {{Islam}}, \citenamefont {{Ma}}, \citenamefont {{Preiss}}, \citenamefont
  {{Tai}}, \citenamefont {{Lukin}}, \citenamefont {{Rispoli}},\ and\
  \citenamefont {{Greiner}}}]{IR15}%
  \BibitemOpen
  \bibfield  {author} {\bibinfo {author} {\bibfnamefont {R.}~\bibnamefont
  {{Islam}}}, \bibinfo {author} {\bibfnamefont {R.}~\bibnamefont {{Ma}}},
  \bibinfo {author} {\bibfnamefont {P.~M.}\ \bibnamefont {{Preiss}}}, \bibinfo
  {author} {\bibfnamefont {M.~E.}\ \bibnamefont {{Tai}}}, \bibinfo {author}
  {\bibfnamefont {A.}~\bibnamefont {{Lukin}}}, \bibinfo {author} {\bibfnamefont
  {M.}~\bibnamefont {{Rispoli}}}, \ and\ \bibinfo {author} {\bibfnamefont
  {M.}~\bibnamefont {{Greiner}}},\ }\href
  {http://www.nature.com/nature/journal/v528/n7580/full/nature15750.html}
  {\bibfield  {journal} {\bibinfo  {journal} {Nature}\ }\textbf {\bibinfo
  {volume} {528}},\ \bibinfo {pages} {77} (\bibinfo {year} {2015})}\BibitemShut
  {NoStop}%
\bibitem [{\citenamefont {Gemelke}\ \emph {et~al.}(2009)\citenamefont
  {Gemelke}, \citenamefont {Zhang}, \citenamefont {Hung},\ and\ \citenamefont
  {Chin}}]{GN09}%
  \BibitemOpen
  \bibfield  {author} {\bibinfo {author} {\bibfnamefont {N.}~\bibnamefont
  {Gemelke}}, \bibinfo {author} {\bibfnamefont {X.}~\bibnamefont {Zhang}},
  \bibinfo {author} {\bibfnamefont {C.-L.}\ \bibnamefont {Hung}}, \ and\
  \bibinfo {author} {\bibfnamefont {C.}~\bibnamefont {Chin}},\ }\href {\doibase
  10.1038/nature08244} {\bibfield  {journal} {\bibinfo  {journal} {Nature}\
  }\textbf {\bibinfo {volume} {460}},\ \bibinfo {pages} {995} (\bibinfo {year}
  {2009})}\BibitemShut {NoStop}%
\bibitem [{\citenamefont {Patil}\ \emph {et~al.}(2014)\citenamefont {Patil},
  \citenamefont {Chakram}, \citenamefont {Aycock},\ and\ \citenamefont
  {Vengalattore}}]{PYS14}%
  \BibitemOpen
  \bibfield  {author} {\bibinfo {author} {\bibfnamefont {Y.~S.}\ \bibnamefont
  {Patil}}, \bibinfo {author} {\bibfnamefont {S.}~\bibnamefont {Chakram}},
  \bibinfo {author} {\bibfnamefont {L.~M.}\ \bibnamefont {Aycock}}, \ and\
  \bibinfo {author} {\bibfnamefont {M.}~\bibnamefont {Vengalattore}},\ }\href
  {\doibase 10.1103/PhysRevA.90.033422} {\bibfield  {journal} {\bibinfo
  {journal} {Phys. Rev. A}\ }\textbf {\bibinfo {volume} {90}},\ \bibinfo
  {pages} {033422} (\bibinfo {year} {2014})}\BibitemShut {NoStop}%
\bibitem [{\citenamefont {Patil}\ \emph {et~al.}(2015)\citenamefont {Patil},
  \citenamefont {Chakram},\ and\ \citenamefont {Vengalattore}}]{PYS142}%
  \BibitemOpen
  \bibfield  {author} {\bibinfo {author} {\bibfnamefont {Y.~S.}\ \bibnamefont
  {Patil}}, \bibinfo {author} {\bibfnamefont {S.}~\bibnamefont {Chakram}}, \
  and\ \bibinfo {author} {\bibfnamefont {M.}~\bibnamefont {Vengalattore}},\
  }\href {\doibase 10.1103/PhysRevLett.115.140402} {\bibfield  {journal}
  {\bibinfo  {journal} {Phys. Rev. Lett.}\ }\textbf {\bibinfo {volume} {115}},\
  \bibinfo {pages} {140402} (\bibinfo {year} {2015})}\BibitemShut {NoStop}%
\bibitem [{\citenamefont {Preiss}\ \emph
  {et~al.}(2015{\natexlab{b}})\citenamefont {Preiss}, \citenamefont {Ma},
  \citenamefont {Tai}, \citenamefont {Simon},\ and\ \citenamefont
  {Greiner}}]{PPM152}%
  \BibitemOpen
  \bibfield  {author} {\bibinfo {author} {\bibfnamefont {P.~M.}\ \bibnamefont
  {Preiss}}, \bibinfo {author} {\bibfnamefont {R.}~\bibnamefont {Ma}}, \bibinfo
  {author} {\bibfnamefont {M.~E.}\ \bibnamefont {Tai}}, \bibinfo {author}
  {\bibfnamefont {J.}~\bibnamefont {Simon}}, \ and\ \bibinfo {author}
  {\bibfnamefont {M.}~\bibnamefont {Greiner}},\ }\href {\doibase
  10.1103/PhysRevA.91.041602} {\bibfield  {journal} {\bibinfo  {journal} {Phys.
  Rev. A}\ }\textbf {\bibinfo {volume} {91}},\ \bibinfo {pages} {041602}
  (\bibinfo {year} {2015}{\natexlab{b}})}\BibitemShut {NoStop}%
\bibitem [{\citenamefont {Ashida}\ and\ \citenamefont {Ueda}(2015)}]{YA15}%
  \BibitemOpen
  \bibfield  {author} {\bibinfo {author} {\bibfnamefont {Y.}~\bibnamefont
  {Ashida}}\ and\ \bibinfo {author} {\bibfnamefont {M.}~\bibnamefont {Ueda}},\
  }\href {\doibase 10.1103/PhysRevLett.115.095301} {\bibfield  {journal}
  {\bibinfo  {journal} {Phys. Rev. Lett.}\ }\textbf {\bibinfo {volume} {115}},\
  \bibinfo {pages} {095301} (\bibinfo {year} {2015})}\BibitemShut {NoStop}%
\bibitem [{\citenamefont {Ashida}\ and\ \citenamefont {Ueda}(2016)}]{YA16}%
  \BibitemOpen
  \bibfield  {author} {\bibinfo {author} {\bibfnamefont {Y.}~\bibnamefont
  {Ashida}}\ and\ \bibinfo {author} {\bibfnamefont {M.}~\bibnamefont {Ueda}},\
  }\href {\doibase 10.1364/OL.41.000072} {\bibfield  {journal} {\bibinfo
  {journal} {Opt. Lett.}\ }\textbf {\bibinfo {volume} {41}},\ \bibinfo {pages}
  {72} (\bibinfo {year} {2016})}\BibitemShut {NoStop}%
\bibitem [{\citenamefont {Mazzucchi}\ \emph {et~al.}(2016)\citenamefont
  {Mazzucchi}, \citenamefont {Kozlowski}, \citenamefont {Caballero-Benitez},
  \citenamefont {Elliott},\ and\ \citenamefont {Mekhov}}]{MG15}%
  \BibitemOpen
  \bibfield  {author} {\bibinfo {author} {\bibfnamefont {G.}~\bibnamefont
  {Mazzucchi}}, \bibinfo {author} {\bibfnamefont {W.}~\bibnamefont
  {Kozlowski}}, \bibinfo {author} {\bibfnamefont {S.~F.}\ \bibnamefont
  {Caballero-Benitez}}, \bibinfo {author} {\bibfnamefont {T.~J.}\ \bibnamefont
  {Elliott}}, \ and\ \bibinfo {author} {\bibfnamefont {I.~B.}\ \bibnamefont
  {Mekhov}},\ }\href {\doibase 10.1103/PhysRevA.93.023632} {\bibfield
  {journal} {\bibinfo  {journal} {Phys. Rev. A}\ }\textbf {\bibinfo {volume}
  {93}},\ \bibinfo {pages} {023632} (\bibinfo {year} {2016})}\BibitemShut
  {NoStop}%
\bibitem [{\citenamefont {Alberti}\ \emph {et~al.}(2016)\citenamefont
  {Alberti}, \citenamefont {Robens}, \citenamefont {Alt}, \citenamefont
  {Brakhane}, \citenamefont {Karski}, \citenamefont {Reimann}, \citenamefont
  {Widera},\ and\ \citenamefont {Meschede}}]{AA15}%
  \BibitemOpen
  \bibfield  {author} {\bibinfo {author} {\bibfnamefont {A.}~\bibnamefont
  {Alberti}}, \bibinfo {author} {\bibfnamefont {C.}~\bibnamefont {Robens}},
  \bibinfo {author} {\bibfnamefont {W.}~\bibnamefont {Alt}}, \bibinfo {author}
  {\bibfnamefont {S.}~\bibnamefont {Brakhane}}, \bibinfo {author}
  {\bibfnamefont {M.}~\bibnamefont {Karski}}, \bibinfo {author} {\bibfnamefont
  {R.}~\bibnamefont {Reimann}}, \bibinfo {author} {\bibfnamefont
  {A.}~\bibnamefont {Widera}}, \ and\ \bibinfo {author} {\bibfnamefont
  {D.}~\bibnamefont {Meschede}},\ }\href
  {http://iopscience.iop.org/article/10.1088/1367-2630/18/5/053010} {\bibfield
  {journal} {\bibinfo  {journal} {New J. Phys.}\ }\textbf {\bibinfo {volume}
  {18}},\ \bibinfo {pages} {053010} (\bibinfo {year} {2016})}\BibitemShut
  {NoStop}%
\bibitem [{\citenamefont {{Wigley}}\ \emph {et~al.}(2016)\citenamefont
  {{Wigley}}, \citenamefont {{Everitt}}, \citenamefont {{Hardman}},
  \citenamefont {{Sooriyabandara}}, \citenamefont {{Perumbil}}, \citenamefont
  {{Close}}, \citenamefont {{Robins}},\ and\ \citenamefont {{Kuhn}}}]{WP16}%
  \BibitemOpen
  \bibfield  {author} {\bibinfo {author} {\bibfnamefont {P.}~\bibnamefont
  {{Wigley}}}, \bibinfo {author} {\bibfnamefont {P.}~\bibnamefont {{Everitt}}},
  \bibinfo {author} {\bibfnamefont {K.}~\bibnamefont {{Hardman}}}, \bibinfo
  {author} {\bibfnamefont {M.}~\bibnamefont {{Sooriyabandara}}}, \bibinfo
  {author} {\bibfnamefont {M.}~\bibnamefont {{Perumbil}}}, \bibinfo {author}
  {\bibfnamefont {J.}~\bibnamefont {{Close}}}, \bibinfo {author} {\bibfnamefont
  {N.}~\bibnamefont {{Robins}}}, \ and\ \bibinfo {author} {\bibfnamefont
  {C.}~\bibnamefont {{Kuhn}}},\ }\href@noop {} {\bibfield  {journal} {\bibinfo
  {journal} {ArXiv e-prints}\ } (\bibinfo {year} {2016})},\ \Eprint
  {http://arxiv.org/abs/1601.04425} {arXiv:1601.04425} \BibitemShut {NoStop}%
\bibitem [{\citenamefont {Bloch}\ \emph {et~al.}(2012)\citenamefont {Bloch},
  \citenamefont {Dalibard},\ and\ \citenamefont {Nascimb{\`e}ne}}]{BI12}%
  \BibitemOpen
  \bibfield  {author} {\bibinfo {author} {\bibfnamefont {I.}~\bibnamefont
  {Bloch}}, \bibinfo {author} {\bibfnamefont {J.}~\bibnamefont {Dalibard}}, \
  and\ \bibinfo {author} {\bibfnamefont {S.}~\bibnamefont {Nascimb{\`e}ne}},\
  }\href {http://www.nature.com/nphys/journal/v8/n4/abs/nphys2259.html}
  {\bibfield  {journal} {\bibinfo  {journal} {Nat. Phys.}\ }\textbf {\bibinfo
  {volume} {8}},\ \bibinfo {pages} {267} (\bibinfo {year} {2012})}\BibitemShut
  {NoStop}%
\bibitem [{\citenamefont {Wiseman}\ and\ \citenamefont
  {Milburn}(1993)}]{WHM93}%
  \BibitemOpen
  \bibfield  {author} {\bibinfo {author} {\bibfnamefont {H.~M.}\ \bibnamefont
  {Wiseman}}\ and\ \bibinfo {author} {\bibfnamefont {G.~J.}\ \bibnamefont
  {Milburn}},\ }\href {\doibase 10.1103/PhysRevLett.70.548} {\bibfield
  {journal} {\bibinfo  {journal} {Phys. Rev. Lett.}\ }\textbf {\bibinfo
  {volume} {70}},\ \bibinfo {pages} {548} (\bibinfo {year} {1993})}\BibitemShut
  {NoStop}%
\bibitem [{\citenamefont {{Sayrin}}\ \emph {et~al.}(2011)\citenamefont
  {{Sayrin}}, \citenamefont {{Dotsenko}}, \citenamefont {{Zhou}}, \citenamefont
  {{Peaudecerf}}, \citenamefont {{Rybarczyk}}, \citenamefont {{Gleyzes}},
  \citenamefont {{Rouchon}}, \citenamefont {{Mirrahimi}}, \citenamefont
  {{Amini}}, \citenamefont {{Brune}}, \citenamefont {{Raimond}},\ and\
  \citenamefont {{Haroche}}}]{SC11}%
  \BibitemOpen
  \bibfield  {author} {\bibinfo {author} {\bibfnamefont {C.}~\bibnamefont
  {{Sayrin}}}, \bibinfo {author} {\bibfnamefont {I.}~\bibnamefont
  {{Dotsenko}}}, \bibinfo {author} {\bibfnamefont {X.}~\bibnamefont {{Zhou}}},
  \bibinfo {author} {\bibfnamefont {B.}~\bibnamefont {{Peaudecerf}}}, \bibinfo
  {author} {\bibfnamefont {T.}~\bibnamefont {{Rybarczyk}}}, \bibinfo {author}
  {\bibfnamefont {S.}~\bibnamefont {{Gleyzes}}}, \bibinfo {author}
  {\bibfnamefont {P.}~\bibnamefont {{Rouchon}}}, \bibinfo {author}
  {\bibfnamefont {M.}~\bibnamefont {{Mirrahimi}}}, \bibinfo {author}
  {\bibfnamefont {H.}~\bibnamefont {{Amini}}}, \bibinfo {author} {\bibfnamefont
  {M.}~\bibnamefont {{Brune}}}, \bibinfo {author} {\bibfnamefont {J.-M.}\
  \bibnamefont {{Raimond}}}, \ and\ \bibinfo {author} {\bibfnamefont
  {S.}~\bibnamefont {{Haroche}}},\ }\href
  {http://www.nature.com/nature/journal/v477/n7362/full/nature10376.html}
  {\bibfield  {journal} {\bibinfo  {journal} {Nature}\ }\textbf {\bibinfo
  {volume} {477}},\ \bibinfo {pages} {73} (\bibinfo {year} {2011})}\BibitemShut
  {NoStop}%
\bibitem [{\citenamefont {Nagy}\ \emph {et~al.}(2010)\citenamefont {Nagy},
  \citenamefont {K\'onya}, \citenamefont {Szirmai},\ and\ \citenamefont
  {Domokos}}]{ND10}%
  \BibitemOpen
  \bibfield  {author} {\bibinfo {author} {\bibfnamefont {D.}~\bibnamefont
  {Nagy}}, \bibinfo {author} {\bibfnamefont {G.}~\bibnamefont {K\'onya}},
  \bibinfo {author} {\bibfnamefont {G.}~\bibnamefont {Szirmai}}, \ and\
  \bibinfo {author} {\bibfnamefont {P.}~\bibnamefont {Domokos}},\ }\href
  {\doibase 10.1103/PhysRevLett.104.130401} {\bibfield  {journal} {\bibinfo
  {journal} {Phys. Rev. Lett.}\ }\textbf {\bibinfo {volume} {104}},\ \bibinfo
  {pages} {130401} (\bibinfo {year} {2010})}\BibitemShut {NoStop}%
\bibitem [{\citenamefont {Brennecke}\ \emph {et~al.}(2013)\citenamefont
  {Brennecke}, \citenamefont {Mottl}, \citenamefont {Baumann}, \citenamefont
  {Landig}, \citenamefont {Donner},\ and\ \citenamefont {Esslinger}}]{BF13}%
  \BibitemOpen
  \bibfield  {author} {\bibinfo {author} {\bibfnamefont {F.}~\bibnamefont
  {Brennecke}}, \bibinfo {author} {\bibfnamefont {R.}~\bibnamefont {Mottl}},
  \bibinfo {author} {\bibfnamefont {K.}~\bibnamefont {Baumann}}, \bibinfo
  {author} {\bibfnamefont {R.}~\bibnamefont {Landig}}, \bibinfo {author}
  {\bibfnamefont {T.}~\bibnamefont {Donner}}, \ and\ \bibinfo {author}
  {\bibfnamefont {T.}~\bibnamefont {Esslinger}},\ }\href {\doibase
  10.1073/pnas.1306993110} {\bibfield  {journal} {\bibinfo  {journal} {Proc.
  Natl. Aca. Sci. USA}\ }\textbf {\bibinfo {volume} {110}},\ \bibinfo {pages}
  {11763} (\bibinfo {year} {2013})}\BibitemShut {NoStop}%
\bibitem [{\citenamefont {Vasin}\ \emph {et~al.}(2015)\citenamefont {Vasin},
  \citenamefont {Ryzhov},\ and\ \citenamefont {Vinokur}}]{MV15}%
  \BibitemOpen
  \bibfield  {author} {\bibinfo {author} {\bibfnamefont {Y.}~\bibnamefont
  {Vasin}}, \bibinfo {author} {\bibfnamefont {Y.}~\bibnamefont {Ryzhov}}, \
  and\ \bibinfo {author} {\bibfnamefont {V.~M.}\ \bibnamefont {Vinokur}},\
  }\href {\doibase 10.1038/srep18600} {\bibfield  {journal} {\bibinfo
  {journal} {Sci. Rep.}\ }\textbf {\bibinfo {volume} {5}},\ \bibinfo {pages}
  {18600} (\bibinfo {year} {2015})}\BibitemShut {NoStop}%
\bibitem [{\citenamefont {Ashida}\ \emph {et~al.}(2016)\citenamefont {Ashida},
  \citenamefont {Furukawa},\ and\ \citenamefont {Ueda}}]{YAc16}%
  \BibitemOpen
  \bibfield  {author} {\bibinfo {author} {\bibfnamefont {Y.}~\bibnamefont
  {Ashida}}, \bibinfo {author} {\bibfnamefont {S.}~\bibnamefont {Furukawa}}, \
  and\ \bibinfo {author} {\bibfnamefont {M.}~\bibnamefont {Ueda}},\ }\href
  {\doibase 10.1103/PhysRevA.94.053615} {\bibfield  {journal} {\bibinfo
  {journal} {Phys. Rev. A}\ }\textbf {\bibinfo {volume} {94}},\ \bibinfo
  {pages} {053615} (\bibinfo {year} {2016})}\BibitemShut {NoStop}%
\bibitem [{\citenamefont {{Ashida}}\ \emph {et~al.}(2016)\citenamefont
  {{Ashida}}, \citenamefont {{Furukawa}},\ and\ \citenamefont
  {{Ueda}}}]{YAc162}%
  \BibitemOpen
  \bibfield  {author} {\bibinfo {author} {\bibfnamefont {Y.}~\bibnamefont
  {{Ashida}}}, \bibinfo {author} {\bibfnamefont {S.}~\bibnamefont
  {{Furukawa}}}, \ and\ \bibinfo {author} {\bibfnamefont {M.}~\bibnamefont
  {{Ueda}}},\ }\href {https://arxiv.org/abs/1611.00396} {\bibfield  {journal}
  {\bibinfo  {journal} {arXiv:1611.00396}\ } (\bibinfo {year}
  {2016})}\BibitemShut {NoStop}%
\bibitem [{\citenamefont {Barchielli}\ and\ \citenamefont
  {Holevo}(1995)}]{BA95}%
  \BibitemOpen
  \bibfield  {author} {\bibinfo {author} {\bibfnamefont {A.}~\bibnamefont
  {Barchielli}}\ and\ \bibinfo {author} {\bibfnamefont {A.}~\bibnamefont
  {Holevo}},\ }\href
  {http://www.sciencedirect.com/science/article/pii/030441499500011U}
  {\bibfield  {journal} {\bibinfo  {journal} {Stoch. Proc. Appl.}\ }\textbf
  {\bibinfo {volume} {58}},\ \bibinfo {pages} {293 } (\bibinfo {year}
  {1995})}\BibitemShut {NoStop}%
\bibitem [{\citenamefont {Barchielli}\ \emph {et~al.}(1998)\citenamefont
  {Barchielli}, \citenamefont {Paganoni},\ and\ \citenamefont {Zucca}}]{BA98}%
  \BibitemOpen
  \bibfield  {author} {\bibinfo {author} {\bibfnamefont {A.}~\bibnamefont
  {Barchielli}}, \bibinfo {author} {\bibfnamefont {A.}~\bibnamefont
  {Paganoni}}, \ and\ \bibinfo {author} {\bibfnamefont {F.}~\bibnamefont
  {Zucca}},\ }\href
  {http://www.sciencedirect.com/science/article/pii/S0304414997000938}
  {\bibfield  {journal} {\bibinfo  {journal} {Stochastic Processes and their
  Applications}\ }\textbf {\bibinfo {volume} {73}},\ \bibinfo {pages} {69 }
  (\bibinfo {year} {1998})}\BibitemShut {NoStop}%
\bibitem [{\citenamefont {Bassi}\ \emph {et~al.}(2013)\citenamefont {Bassi},
  \citenamefont {Lochan}, \citenamefont {Satin}, \citenamefont {Singh},\ and\
  \citenamefont {Ulbricht}}]{BA13}%
  \BibitemOpen
  \bibfield  {author} {\bibinfo {author} {\bibfnamefont {A.}~\bibnamefont
  {Bassi}}, \bibinfo {author} {\bibfnamefont {K.}~\bibnamefont {Lochan}},
  \bibinfo {author} {\bibfnamefont {S.}~\bibnamefont {Satin}}, \bibinfo
  {author} {\bibfnamefont {T.~P.}\ \bibnamefont {Singh}}, \ and\ \bibinfo
  {author} {\bibfnamefont {H.}~\bibnamefont {Ulbricht}},\ }\href {\doibase
  10.1103/RevModPhys.85.471} {\bibfield  {journal} {\bibinfo  {journal} {Rev.
  Mod. Phys.}\ }\textbf {\bibinfo {volume} {85}},\ \bibinfo {pages} {471}
  (\bibinfo {year} {2013})}\BibitemShut {NoStop}%
\bibitem [{\citenamefont {Pichler}\ \emph {et~al.}(2010)\citenamefont
  {Pichler}, \citenamefont {Daley},\ and\ \citenamefont {Zoller}}]{PH10}%
  \BibitemOpen
  \bibfield  {author} {\bibinfo {author} {\bibfnamefont {H.}~\bibnamefont
  {Pichler}}, \bibinfo {author} {\bibfnamefont {A.~J.}\ \bibnamefont {Daley}},
  \ and\ \bibinfo {author} {\bibfnamefont {P.}~\bibnamefont {Zoller}},\ }\href
  {\doibase 10.1103/PhysRevA.82.063605} {\bibfield  {journal} {\bibinfo
  {journal} {Phys. Rev. A}\ }\textbf {\bibinfo {volume} {82}},\ \bibinfo
  {pages} {063605} (\bibinfo {year} {2010})}\BibitemShut {NoStop}%
\bibitem [{\citenamefont {Ke\ss{}ler}\ \emph {et~al.}(2012)\citenamefont
  {Ke\ss{}ler}, \citenamefont {Holzner}, \citenamefont {McCulloch},
  \citenamefont {von Delft},\ and\ \citenamefont {Marquardt}}]{SK12}%
  \BibitemOpen
  \bibfield  {author} {\bibinfo {author} {\bibfnamefont {S.}~\bibnamefont
  {Ke\ss{}ler}}, \bibinfo {author} {\bibfnamefont {A.}~\bibnamefont {Holzner}},
  \bibinfo {author} {\bibfnamefont {I.~P.}\ \bibnamefont {McCulloch}}, \bibinfo
  {author} {\bibfnamefont {J.}~\bibnamefont {von Delft}}, \ and\ \bibinfo
  {author} {\bibfnamefont {F.}~\bibnamefont {Marquardt}},\ }\href {\doibase
  10.1103/PhysRevA.85.011605} {\bibfield  {journal} {\bibinfo  {journal} {Phys.
  Rev. A}\ }\textbf {\bibinfo {volume} {85}},\ \bibinfo {pages} {011605}
  (\bibinfo {year} {2012})}\BibitemShut {NoStop}%
\bibitem [{\citenamefont {Poletti}\ \emph {et~al.}(2012)\citenamefont
  {Poletti}, \citenamefont {Bernier}, \citenamefont {Georges},\ and\
  \citenamefont {Kollath}}]{PD12}%
  \BibitemOpen
  \bibfield  {author} {\bibinfo {author} {\bibfnamefont {D.}~\bibnamefont
  {Poletti}}, \bibinfo {author} {\bibfnamefont {J.-S.}\ \bibnamefont
  {Bernier}}, \bibinfo {author} {\bibfnamefont {A.}~\bibnamefont {Georges}}, \
  and\ \bibinfo {author} {\bibfnamefont {C.}~\bibnamefont {Kollath}},\ }\href
  {\doibase 10.1103/PhysRevLett.109.045302} {\bibfield  {journal} {\bibinfo
  {journal} {Phys. Rev. Lett.}\ }\textbf {\bibinfo {volume} {109}},\ \bibinfo
  {pages} {045302} (\bibinfo {year} {2012})}\BibitemShut {NoStop}%
\bibitem [{\citenamefont {Poletti}\ \emph {et~al.}(2013)\citenamefont
  {Poletti}, \citenamefont {Barmettler}, \citenamefont {Georges},\ and\
  \citenamefont {Kollath}}]{PD13}%
  \BibitemOpen
  \bibfield  {author} {\bibinfo {author} {\bibfnamefont {D.}~\bibnamefont
  {Poletti}}, \bibinfo {author} {\bibfnamefont {P.}~\bibnamefont {Barmettler}},
  \bibinfo {author} {\bibfnamefont {A.}~\bibnamefont {Georges}}, \ and\
  \bibinfo {author} {\bibfnamefont {C.}~\bibnamefont {Kollath}},\ }\href
  {\doibase 10.1103/PhysRevLett.111.195301} {\bibfield  {journal} {\bibinfo
  {journal} {Phys. Rev. Lett.}\ }\textbf {\bibinfo {volume} {111}},\ \bibinfo
  {pages} {195301} (\bibinfo {year} {2013})}\BibitemShut {NoStop}%
\bibitem [{\citenamefont {Yanay}\ and\ \citenamefont {Mueller}(2014)}]{YY14}%
  \BibitemOpen
  \bibfield  {author} {\bibinfo {author} {\bibfnamefont {Y.}~\bibnamefont
  {Yanay}}\ and\ \bibinfo {author} {\bibfnamefont {E.~J.}\ \bibnamefont
  {Mueller}},\ }\href {\doibase 10.1103/PhysRevA.90.023611} {\bibfield
  {journal} {\bibinfo  {journal} {Phys. Rev. A}\ }\textbf {\bibinfo {volume}
  {90}},\ \bibinfo {pages} {023611} (\bibinfo {year} {2014})}\BibitemShut
  {NoStop}%
\bibitem [{\citenamefont {Caves}\ and\ \citenamefont {Milburn}(1987)}]{CC87}%
  \BibitemOpen
  \bibfield  {author} {\bibinfo {author} {\bibfnamefont {C.~M.}\ \bibnamefont
  {Caves}}\ and\ \bibinfo {author} {\bibfnamefont {G.~J.}\ \bibnamefont
  {Milburn}},\ }\href {\doibase 10.1103/PhysRevA.36.5543} {\bibfield  {journal}
  {\bibinfo  {journal} {Phys. Rev. A}\ }\textbf {\bibinfo {volume} {36}},\
  \bibinfo {pages} {5543} (\bibinfo {year} {1987})}\BibitemShut {NoStop}%
\bibitem [{\citenamefont {Di\'{o}si}(1988{\natexlab{a}})}]{DL88}%
  \BibitemOpen
  \bibfield  {author} {\bibinfo {author} {\bibfnamefont {L.}~\bibnamefont
  {Di\'{o}si}},\ }\href {\doibase
  http://dx.doi.org/10.1016/0375-9601(88)90309-X} {\bibfield  {journal}
  {\bibinfo  {journal} {Phys. Lett. A}\ }\textbf {\bibinfo {volume} {129}},\
  \bibinfo {pages} {419 } (\bibinfo {year} {1988}{\natexlab{a}})}\BibitemShut
  {NoStop}%
\bibitem [{\citenamefont {Di\'{o}si}(1988{\natexlab{b}})}]{DL882}%
  \BibitemOpen
  \bibfield  {author} {\bibinfo {author} {\bibfnamefont {L.}~\bibnamefont
  {Di\'{o}si}},\ }\href {\doibase
  http://dx.doi.org/10.1016/0375-9601(88)90555-5} {\bibfield  {journal}
  {\bibinfo  {journal} {Phys. Lett. A}\ }\textbf {\bibinfo {volume} {132}},\
  \bibinfo {pages} {233 } (\bibinfo {year} {1988}{\natexlab{b}})}\BibitemShut
  {NoStop}%
\bibitem [{\citenamefont {Belavkin}\ and\ \citenamefont
  {Staszewski}(1992)}]{BVP92}%
  \BibitemOpen
  \bibfield  {author} {\bibinfo {author} {\bibfnamefont {V.~P.}\ \bibnamefont
  {Belavkin}}\ and\ \bibinfo {author} {\bibfnamefont {P.}~\bibnamefont
  {Staszewski}},\ }\href {\doibase 10.1103/PhysRevA.45.1347} {\bibfield
  {journal} {\bibinfo  {journal} {Phys. Rev. A}\ }\textbf {\bibinfo {volume}
  {45}},\ \bibinfo {pages} {1347} (\bibinfo {year} {1992})}\BibitemShut
  {NoStop}%
\bibitem [{\citenamefont {Gagen}\ \emph {et~al.}(1993)\citenamefont {Gagen},
  \citenamefont {Wiseman},\ and\ \citenamefont {Milburn}}]{GMJ93}%
  \BibitemOpen
  \bibfield  {author} {\bibinfo {author} {\bibfnamefont {M.~J.}\ \bibnamefont
  {Gagen}}, \bibinfo {author} {\bibfnamefont {H.~M.}\ \bibnamefont {Wiseman}},
  \ and\ \bibinfo {author} {\bibfnamefont {G.~J.}\ \bibnamefont {Milburn}},\
  }\href {\doibase 10.1103/PhysRevA.48.132} {\bibfield  {journal} {\bibinfo
  {journal} {Phys. Rev. A}\ }\textbf {\bibinfo {volume} {48}},\ \bibinfo
  {pages} {132} (\bibinfo {year} {1993})}\BibitemShut {NoStop}%
\bibitem [{\citenamefont {Gullans}\ \emph {et~al.}(2012)\citenamefont
  {Gullans}, \citenamefont {Tiecke}, \citenamefont {Chang}, \citenamefont
  {Feist}, \citenamefont {Thompson}, \citenamefont {Cirac}, \citenamefont
  {Zoller},\ and\ \citenamefont {Lukin}}]{GM12}%
  \BibitemOpen
  \bibfield  {author} {\bibinfo {author} {\bibfnamefont {M.}~\bibnamefont
  {Gullans}}, \bibinfo {author} {\bibfnamefont {T.~G.}\ \bibnamefont {Tiecke}},
  \bibinfo {author} {\bibfnamefont {D.~E.}\ \bibnamefont {Chang}}, \bibinfo
  {author} {\bibfnamefont {J.}~\bibnamefont {Feist}}, \bibinfo {author}
  {\bibfnamefont {J.~D.}\ \bibnamefont {Thompson}}, \bibinfo {author}
  {\bibfnamefont {J.~I.}\ \bibnamefont {Cirac}}, \bibinfo {author}
  {\bibfnamefont {P.}~\bibnamefont {Zoller}}, \ and\ \bibinfo {author}
  {\bibfnamefont {M.~D.}\ \bibnamefont {Lukin}},\ }\href {\doibase
  10.1103/PhysRevLett.109.235309} {\bibfield  {journal} {\bibinfo  {journal}
  {Phys. Rev. Lett.}\ }\textbf {\bibinfo {volume} {109}},\ \bibinfo {pages}
  {235309} (\bibinfo {year} {2012})}\BibitemShut {NoStop}%
\bibitem [{\citenamefont {Romero-Isart}\ \emph {et~al.}(2013)\citenamefont
  {Romero-Isart}, \citenamefont {Navau}, \citenamefont {Sanchez}, \citenamefont
  {Zoller},\ and\ \citenamefont {Cirac}}]{RIO13}%
  \BibitemOpen
  \bibfield  {author} {\bibinfo {author} {\bibfnamefont {O.}~\bibnamefont
  {Romero-Isart}}, \bibinfo {author} {\bibfnamefont {C.}~\bibnamefont {Navau}},
  \bibinfo {author} {\bibfnamefont {A.}~\bibnamefont {Sanchez}}, \bibinfo
  {author} {\bibfnamefont {P.}~\bibnamefont {Zoller}}, \ and\ \bibinfo {author}
  {\bibfnamefont {J.~I.}\ \bibnamefont {Cirac}},\ }\href {\doibase
  10.1103/PhysRevLett.111.145304} {\bibfield  {journal} {\bibinfo  {journal}
  {Phys. Rev. Lett.}\ }\textbf {\bibinfo {volume} {111}},\ \bibinfo {pages}
  {145304} (\bibinfo {year} {2013})}\BibitemShut {NoStop}%
\bibitem [{\citenamefont {Gonz{\'a}lez-Tudela}\ \emph
  {et~al.}(2015)\citenamefont {Gonz{\'a}lez-Tudela}, \citenamefont {Hung},
  \citenamefont {Chang}, \citenamefont {Cirac},\ and\ \citenamefont
  {Kimble}}]{GT15}%
  \BibitemOpen
  \bibfield  {author} {\bibinfo {author} {\bibfnamefont {A.}~\bibnamefont
  {Gonz{\'a}lez-Tudela}}, \bibinfo {author} {\bibfnamefont {C.-L.}\
  \bibnamefont {Hung}}, \bibinfo {author} {\bibfnamefont {D.~E.}\ \bibnamefont
  {Chang}}, \bibinfo {author} {\bibfnamefont {I.~J.}\ \bibnamefont {Cirac}}, \
  and\ \bibinfo {author} {\bibfnamefont {J.~H.}\ \bibnamefont {Kimble}},\
  }\href
  {http://www.nature.com/nphoton/journal/v9/n5/full/nphoton.2015.54.html}
  {\bibfield  {journal} {\bibinfo  {journal} {Nat. Photon.}\ }\textbf {\bibinfo
  {volume} {9}},\ \bibinfo {pages} {320} (\bibinfo {year} {2015})}\BibitemShut
  {NoStop}%
\bibitem [{\citenamefont {Nascimbene}\ \emph {et~al.}(2015)\citenamefont
  {Nascimbene}, \citenamefont {Goldman}, \citenamefont {Cooper},\ and\
  \citenamefont {Dalibard}}]{SN15}%
  \BibitemOpen
  \bibfield  {author} {\bibinfo {author} {\bibfnamefont {S.}~\bibnamefont
  {Nascimbene}}, \bibinfo {author} {\bibfnamefont {N.}~\bibnamefont {Goldman}},
  \bibinfo {author} {\bibfnamefont {N.~R.}\ \bibnamefont {Cooper}}, \ and\
  \bibinfo {author} {\bibfnamefont {J.}~\bibnamefont {Dalibard}},\ }\href
  {\doibase 10.1103/PhysRevLett.115.140401} {\bibfield  {journal} {\bibinfo
  {journal} {Phys. Rev. Lett.}\ }\textbf {\bibinfo {volume} {115}},\ \bibinfo
  {pages} {140401} (\bibinfo {year} {2015})}\BibitemShut {NoStop}%
\bibitem [{\citenamefont {Breuer}\ and\ \citenamefont
  {Petruccione}(2002)}]{HPB02}%
  \BibitemOpen
  \bibfield  {author} {\bibinfo {author} {\bibfnamefont {H.-P.}\ \bibnamefont
  {Breuer}}\ and\ \bibinfo {author} {\bibfnamefont {F.}~\bibnamefont
  {Petruccione}},\ }\href@noop {} {\emph {\bibinfo {title} {The Theory of Open
  Quantum Systems}}}\ (\bibinfo  {publisher} {Oxford University Press},\
  \bibinfo {year} {2002})\BibitemShut {NoStop}%
\bibitem [{\citenamefont {Barchielli}\ and\ \citenamefont
  {Gregoratti}(2012)}]{BA12}%
  \BibitemOpen
  \bibfield  {author} {\bibinfo {author} {\bibfnamefont {A.}~\bibnamefont
  {Barchielli}}\ and\ \bibinfo {author} {\bibfnamefont {M.}~\bibnamefont
  {Gregoratti}},\ }\href {\doibase 10.1098/rsta.2011.0515} {\bibfield
  {journal} {\bibinfo  {journal} {Phil. Trans. R. Soc. A}\ }\textbf {\bibinfo
  {volume} {370}},\ \bibinfo {pages} {5364} (\bibinfo {year}
  {2012})}\BibitemShut {NoStop}%
\bibitem [{\citenamefont {Di\'osi}(1989)}]{DL89}%
  \BibitemOpen
  \bibfield  {author} {\bibinfo {author} {\bibfnamefont {L.}~\bibnamefont
  {Di\'osi}},\ }\href {\doibase 10.1103/PhysRevA.40.1165} {\bibfield  {journal}
  {\bibinfo  {journal} {Phys. Rev. A}\ }\textbf {\bibinfo {volume} {40}},\
  \bibinfo {pages} {1165} (\bibinfo {year} {1989})}\BibitemShut {NoStop}%
\bibitem [{\citenamefont {Douglas}\ and\ \citenamefont
  {Burnett}(2010)}]{DJS10}%
  \BibitemOpen
  \bibfield  {author} {\bibinfo {author} {\bibfnamefont {J.~S.}\ \bibnamefont
  {Douglas}}\ and\ \bibinfo {author} {\bibfnamefont {K.}~\bibnamefont
  {Burnett}},\ }\href {\doibase 10.1103/PhysRevA.82.033434} {\bibfield
  {journal} {\bibinfo  {journal} {Phys. Rev. A}\ }\textbf {\bibinfo {volume}
  {82}},\ \bibinfo {pages} {033434} (\bibinfo {year} {2010})}\BibitemShut
  {NoStop}%
\bibitem [{\citenamefont {Beige}\ \emph {et~al.}(2000)\citenamefont {Beige},
  \citenamefont {Braun}, \citenamefont {Tregenna},\ and\ \citenamefont
  {Knight}}]{BA00}%
  \BibitemOpen
  \bibfield  {author} {\bibinfo {author} {\bibfnamefont {A.}~\bibnamefont
  {Beige}}, \bibinfo {author} {\bibfnamefont {D.}~\bibnamefont {Braun}},
  \bibinfo {author} {\bibfnamefont {B.}~\bibnamefont {Tregenna}}, \ and\
  \bibinfo {author} {\bibfnamefont {P.~L.}\ \bibnamefont {Knight}},\ }\href
  {\doibase 10.1103/PhysRevLett.85.1762} {\bibfield  {journal} {\bibinfo
  {journal} {Phys. Rev. Lett.}\ }\textbf {\bibinfo {volume} {85}},\ \bibinfo
  {pages} {1762} (\bibinfo {year} {2000})}\BibitemShut {NoStop}%
\bibitem [{\citenamefont {Facchi}\ and\ \citenamefont {Pascazio}(2002)}]{FP02}%
  \BibitemOpen
  \bibfield  {author} {\bibinfo {author} {\bibfnamefont {P.}~\bibnamefont
  {Facchi}}\ and\ \bibinfo {author} {\bibfnamefont {S.}~\bibnamefont
  {Pascazio}},\ }\href {\doibase 10.1103/PhysRevLett.89.080401} {\bibfield
  {journal} {\bibinfo  {journal} {Phys. Rev. Lett.}\ }\textbf {\bibinfo
  {volume} {89}},\ \bibinfo {pages} {080401} (\bibinfo {year}
  {2002})}\BibitemShut {NoStop}%
\bibitem [{\citenamefont {Ghirardi}\ \emph {et~al.}(1986)\citenamefont
  {Ghirardi}, \citenamefont {Rimini},\ and\ \citenamefont {Weber}}]{GGC86}%
  \BibitemOpen
  \bibfield  {author} {\bibinfo {author} {\bibfnamefont {G.~C.}\ \bibnamefont
  {Ghirardi}}, \bibinfo {author} {\bibfnamefont {A.}~\bibnamefont {Rimini}}, \
  and\ \bibinfo {author} {\bibfnamefont {T.}~\bibnamefont {Weber}},\ }\href
  {\doibase 10.1103/PhysRevD.34.470} {\bibfield  {journal} {\bibinfo  {journal}
  {Phys. Rev. D}\ }\textbf {\bibinfo {volume} {34}},\ \bibinfo {pages} {470}
  (\bibinfo {year} {1986})}\BibitemShut {NoStop}%
\bibitem [{\citenamefont {Ghirardi}\ \emph {et~al.}(1990)\citenamefont
  {Ghirardi}, \citenamefont {Pearle},\ and\ \citenamefont {Rimini}}]{GGC90}%
  \BibitemOpen
  \bibfield  {author} {\bibinfo {author} {\bibfnamefont {G.~C.}\ \bibnamefont
  {Ghirardi}}, \bibinfo {author} {\bibfnamefont {P.}~\bibnamefont {Pearle}}, \
  and\ \bibinfo {author} {\bibfnamefont {A.}~\bibnamefont {Rimini}},\ }\href
  {\doibase 10.1103/PhysRevA.42.78} {\bibfield  {journal} {\bibinfo  {journal}
  {Phys. Rev. A}\ }\textbf {\bibinfo {volume} {42}},\ \bibinfo {pages} {78}
  (\bibinfo {year} {1990})}\BibitemShut {NoStop}%
\bibitem [{\citenamefont {Lahini}\ \emph {et~al.}(2012)\citenamefont {Lahini},
  \citenamefont {Verbin}, \citenamefont {Huber}, \citenamefont {Bromberg},
  \citenamefont {Pugatch},\ and\ \citenamefont {Silberberg}}]{LY12}%
  \BibitemOpen
  \bibfield  {author} {\bibinfo {author} {\bibfnamefont {Y.}~\bibnamefont
  {Lahini}}, \bibinfo {author} {\bibfnamefont {M.}~\bibnamefont {Verbin}},
  \bibinfo {author} {\bibfnamefont {S.~D.}\ \bibnamefont {Huber}}, \bibinfo
  {author} {\bibfnamefont {Y.}~\bibnamefont {Bromberg}}, \bibinfo {author}
  {\bibfnamefont {R.}~\bibnamefont {Pugatch}}, \ and\ \bibinfo {author}
  {\bibfnamefont {Y.}~\bibnamefont {Silberberg}},\ }\href {\doibase
  10.1103/PhysRevA.86.011603} {\bibfield  {journal} {\bibinfo  {journal} {Phys.
  Rev. A}\ }\textbf {\bibinfo {volume} {86}},\ \bibinfo {pages} {011603}
  (\bibinfo {year} {2012})}\BibitemShut {NoStop}%
\bibitem [{\citenamefont {Grifoni}\ and\ \citenamefont {H{\"
  a}nggi}(1998)}]{MG98}%
  \BibitemOpen
  \bibfield  {author} {\bibinfo {author} {\bibfnamefont {M.}~\bibnamefont
  {Grifoni}}\ and\ \bibinfo {author} {\bibfnamefont {P.}~\bibnamefont {H{\"
  a}nggi}},\ }\href {\doibase http://dx.doi.org/10.1016/S0370-1573(98)00022-2}
  {\bibfield  {journal} {\bibinfo  {journal} {Phys. Rep.}\ }\textbf {\bibinfo
  {volume} {304}},\ \bibinfo {pages} {229 } (\bibinfo {year}
  {1998})}\BibitemShut {NoStop}%
\bibitem [{\citenamefont {Schwarzer}\ and\ \citenamefont {Haken}(1972)}]{ES72}%
  \BibitemOpen
  \bibfield  {author} {\bibinfo {author} {\bibfnamefont {E.}~\bibnamefont
  {Schwarzer}}\ and\ \bibinfo {author} {\bibfnamefont {H.}~\bibnamefont
  {Haken}},\ }\href {\doibase http://dx.doi.org/10.1016/0375-9601(72)90439-2}
  {\bibfield  {journal} {\bibinfo  {journal} {Phys. Lett. A}\ }\textbf
  {\bibinfo {volume} {42}},\ \bibinfo {pages} {317 } (\bibinfo {year}
  {1972})}\BibitemShut {NoStop}%
\bibitem [{\citenamefont {Brun}\ \emph {et~al.}(2003)\citenamefont {Brun},
  \citenamefont {Carteret},\ and\ \citenamefont {Ambainis}}]{BTA03}%
  \BibitemOpen
  \bibfield  {author} {\bibinfo {author} {\bibfnamefont {T.~A.}\ \bibnamefont
  {Brun}}, \bibinfo {author} {\bibfnamefont {H.~A.}\ \bibnamefont {Carteret}},
  \ and\ \bibinfo {author} {\bibfnamefont {A.}~\bibnamefont {Ambainis}},\
  }\href {\doibase 10.1103/PhysRevLett.91.130602} {\bibfield  {journal}
  {\bibinfo  {journal} {Phys. Rev. Lett.}\ }\textbf {\bibinfo {volume} {91}},\
  \bibinfo {pages} {130602} (\bibinfo {year} {2003})}\BibitemShut {NoStop}%
\bibitem [{\citenamefont {Hoyer}\ \emph {et~al.}(2010)\citenamefont {Hoyer},
  \citenamefont {Sarovar},\ and\ \citenamefont {Whaley}}]{SH10}%
  \BibitemOpen
  \bibfield  {author} {\bibinfo {author} {\bibfnamefont {S.}~\bibnamefont
  {Hoyer}}, \bibinfo {author} {\bibfnamefont {M.}~\bibnamefont {Sarovar}}, \
  and\ \bibinfo {author} {\bibfnamefont {K.~B.}\ \bibnamefont {Whaley}},\
  }\href {http://iopscience.iop.org/article/10.1088/1367-2630/12/6/065041/meta}
  {\bibfield  {journal} {\bibinfo  {journal} {New J. Phys.}\ }\textbf {\bibinfo
  {volume} {12}},\ \bibinfo {pages} {065041} (\bibinfo {year}
  {2010})}\BibitemShut {NoStop}%
\bibitem [{\citenamefont {Schreiber}\ \emph {et~al.}(2011)\citenamefont
  {Schreiber}, \citenamefont {Cassemiro}, \citenamefont
  {Poto\ifmmode~\check{c}\else \v{c}\fi{}ek}, \citenamefont {G\'abris},
  \citenamefont {Jex},\ and\ \citenamefont {Silberhorn}}]{SA11}%
  \BibitemOpen
  \bibfield  {author} {\bibinfo {author} {\bibfnamefont {A.}~\bibnamefont
  {Schreiber}}, \bibinfo {author} {\bibfnamefont {K.~N.}\ \bibnamefont
  {Cassemiro}}, \bibinfo {author} {\bibfnamefont {V.}~\bibnamefont
  {Poto\ifmmode~\check{c}\else \v{c}\fi{}ek}}, \bibinfo {author} {\bibfnamefont
  {A.}~\bibnamefont {G\'abris}}, \bibinfo {author} {\bibfnamefont
  {I.}~\bibnamefont {Jex}}, \ and\ \bibinfo {author} {\bibfnamefont
  {C.}~\bibnamefont {Silberhorn}},\ }\href {\doibase
  10.1103/PhysRevLett.106.180403} {\bibfield  {journal} {\bibinfo  {journal}
  {Phys. Rev. Lett.}\ }\textbf {\bibinfo {volume} {106}},\ \bibinfo {pages}
  {180403} (\bibinfo {year} {2011})}\BibitemShut {NoStop}%
\bibitem [{\citenamefont {Jacobs}\ and\ \citenamefont {Steck}(2006)}]{KJ06}%
  \BibitemOpen
  \bibfield  {author} {\bibinfo {author} {\bibfnamefont {K.}~\bibnamefont
  {Jacobs}}\ and\ \bibinfo {author} {\bibfnamefont {D.~A.}\ \bibnamefont
  {Steck}},\ }\href {\doibase 10.1080/00107510601101934} {\bibfield  {journal}
  {\bibinfo  {journal} {Contemp. Phys.}\ }\textbf {\bibinfo {volume} {47}},\
  \bibinfo {pages} {279} (\bibinfo {year} {2006})}\BibitemShut {NoStop}%
\bibitem [{\citenamefont {Eschner}\ \emph {et~al.}(2003)\citenamefont
  {Eschner}, \citenamefont {Morigi}, \citenamefont {Schmidt-Kaler},\ and\
  \citenamefont {Blatt}}]{JE03}%
  \BibitemOpen
  \bibfield  {author} {\bibinfo {author} {\bibfnamefont {J.}~\bibnamefont
  {Eschner}}, \bibinfo {author} {\bibfnamefont {G.}~\bibnamefont {Morigi}},
  \bibinfo {author} {\bibfnamefont {F.}~\bibnamefont {Schmidt-Kaler}}, \ and\
  \bibinfo {author} {\bibfnamefont {R.}~\bibnamefont {Blatt}},\ }\href
  {\doibase 10.1364/JOSAB.20.001003} {\bibfield  {journal} {\bibinfo  {journal}
  {J. Opt. Soc. Am. B}\ }\textbf {\bibinfo {volume} {20}},\ \bibinfo {pages}
  {1003} (\bibinfo {year} {2003})}\BibitemShut {NoStop}%
\bibitem [{\citenamefont {Weitenberg}\ \emph
  {et~al.}(2011{\natexlab{b}})\citenamefont {Weitenberg}, \citenamefont
  {Schau\ss{}}, \citenamefont {Fukuhara}, \citenamefont {Cheneau},
  \citenamefont {Endres}, \citenamefont {Bloch},\ and\ \citenamefont
  {Kuhr}}]{WC112}%
  \BibitemOpen
  \bibfield  {author} {\bibinfo {author} {\bibfnamefont {C.}~\bibnamefont
  {Weitenberg}}, \bibinfo {author} {\bibfnamefont {P.}~\bibnamefont
  {Schau\ss{}}}, \bibinfo {author} {\bibfnamefont {T.}~\bibnamefont
  {Fukuhara}}, \bibinfo {author} {\bibfnamefont {M.}~\bibnamefont {Cheneau}},
  \bibinfo {author} {\bibfnamefont {M.}~\bibnamefont {Endres}}, \bibinfo
  {author} {\bibfnamefont {I.}~\bibnamefont {Bloch}}, \ and\ \bibinfo {author}
  {\bibfnamefont {S.}~\bibnamefont {Kuhr}},\ }\href {\doibase
  10.1103/PhysRevLett.106.215301} {\bibfield  {journal} {\bibinfo  {journal}
  {Phys. Rev. Lett.}\ }\textbf {\bibinfo {volume} {106}},\ \bibinfo {pages}
  {215301} (\bibinfo {year} {2011}{\natexlab{b}})}\BibitemShut {NoStop}%
\end{thebibliography}%

\end{document}